\documentclass[12pt,journal,draftclsnofoot,onecolumn,letterpaper]{IEEEtran}

\usepackage{amsmath, amsthm}
\usepackage{eepic, epic,color}
\usepackage{times}
\usepackage[latin1]{inputenc}
\usepackage{amsfonts}
\usepackage{psfrag}
\usepackage{amsbsy}
\usepackage{amssymb}
\usepackage[mathscr]{eucal}
\usepackage{latexsym}
\usepackage[T1]{fontenc}
\usepackage{amsmath}
\usepackage{comment}
\usepackage[dvips]{graphicx}
\usepackage{bm}
\usepackage{array}
\usepackage{verbatim,mathrsfs, cite}
\usepackage{geometry}
\usepackage[options]{mcode}
\usepackage{algorithmic}
\usepackage{algorithm}
\usepackage{multirow}

\newcounter{lecnum}
\newtheorem{defn}{Definition}
\newtheorem{theorem}{Theorem}
\newtheorem{lemma}[theorem]{Lemma}
\newtheorem{proposition}[theorem]{Proposition}

\newtheorem{corollary}[theorem]{Corollary}

\newcommand{\figw}{0.95\columnwidth}
\newcommand{\argmax}{\operatornamewithlimits{argmax}}
\newcommand{\argmin}{\operatornamewithlimits{argmin}}

\def\B(#1){\hbox{\boldmath$#1$}}
\def\C(#1){{\cal #1}}

\graphicspath{{./Figure/}}

\begin{document}

\title{\LARGE Designing Information Revelation and Intervention with an Application to Flow Control}

\author{\authorblockN{{\bf Luca Canzian}$^\diamond$, {\bf Yuanzhang Xiao}$^\S$, {\bf William Zame}$^+$, {\bf Michele Zorzi}$^\diamond$, \\{\bf Mihaela van der Schaar}$^\S$}\\
{$^\diamond$DEI, University of Padova, via Gradenigo 6/B, 35131 Padova, Italy
%, \{canzian,zorzi\}@dei.unipd.it
\\
$^\S$Department of Electrical Engineering, UCLA, Los Angeles CA 90095, USA
%, \{yxiao,mihaela\}@ee.ucla.edu
\\
$^+$Department of Economics, UCLA, Los Angeles CA 90095, USA
%, zame@econ.ucla.edu
}
}

\maketitle

\begin{abstract}
There are many familiar situations in which a manager seeks to design a system in which users share a resource, but outcomes depend on the information held and actions taken by users.  
If communication is possible, the manager can ask users to report their private information and then, using this information, instruct them on what actions they should take.   
If the users are compliant, this reduces the manager's optimization problem to a well-studied problem of optimal control. 
However, if the users are self-interested and not compliant, the problem is much more complicated: when asked to report their private information, the users might lie; upon receiving instructions, the users might disobey.  
Here we ask whether the manager can design the system to get around both of these difficulties.  
To do so, the manager must provide for the users the incentives to report truthfully and to follow the instructions, despite the fact that the users are self-interested. 
For a class of environments that includes many resource allocation games in communication networks, we provide tools for the manager to design an efficient system. 
In addition to reports and recommendations, the design we employ allows the manager to intervene in the system after the users take actions.
In an abstracted environment, we find conditions under which the manager can achieve the same outcome it could if users were compliant, and conditions under which it does not.
We then apply our framework and results to design a flow control management system. 

%There are many situations in which a manager seeks to design a system that will be used by others and seeks to maximize some social or private objective that depends on information held and actions taken by the others.
%If the users are compliant, the manager can obtain their private information by asking them to report it and then, using this information, instructs them on how to behave. However, if the users are self-interested and hence might not be compliant, the problem is much more complicated. First, when asked to report their private information, the users might lie strategically; second, upon receiving instructions, the users might not follow the instructions. Here we ask whether the manager can design the system to get around both of these problems. On the one hand, the manager must provide the users incentives to report truthfully, on the other hand the manager must provide incentives for the users to follow instructions, despite the fact that the users are self-interested. For a class of environments that includes resource allocation games in communication networks, we provide tools for the manager to design an efficient system. The design we employ has two features: 1) the manager asks the users to report their private information and, based on these reports, recommends actions; 2) following the reports of the users, recommendations of the manager and subsequent actions of the users, the manager may in addition take an action of its own. Finally, we apply the proposed framework to design a flow control management system. 
\end{abstract}


\section{Introduction}\label{sec:intro}

There are many situations in which a manager seeks to design a system for users to share a resource, optimizing it according to some given benevolent or selfish criterion. 
If the manager has full information and users cannot act independently of the manager, the manager's problem is one of optimal control and is well-studied. 
If the users have information the manager does not have and act independently of the manager, but communication between the users and the manager is possible and users are compliant, the manager's problem is only slightly more complicated: the  manager can simply ask the users to report their private information and then provide instructions on how it wishes them to behave. 
Because the users are compliant, they will report truthfully and obey instructions, so, whatever the manager's objective, this again reduces to a known problem in optimal control. 
However, if the users are \emph{self-interested} and \emph{strategic}, two difficulties arise.
The first is that the users might lie about their private information -- if it is in their individual interests to do so; the second is that the users might disobey the instructions of the manager --  if it is in their individual interests to do so. 
The manager's problem in this setting is to design a system to maximize its objective function, given the self-interested and strategic nature of the users. 
A case of particular interest is that of a benevolent manager, who seeks to allocate resources efficiently or fairly according to some measure of social welfare. 
Efficient resource allocation is crucial to make the system accessible to many users and provide each of them with good service. 
However, the problem faced by a benevolent manager may be no easier than the problem faced by a selfish manager, who maximizes some measure of its own personal welfare, because the strategic interests of the individual users will be different from the interests of the group of users as a whole, and hence may still lead individual users to lie and to disobey.\footnote{Even in the absence of private information, the strategic interests of the individual users usually lead to the over-use of resources and to substantial inefficiencies \cite{802, slotAloha}.}

In the  economics literature, such problems are formalized in terms of \emph{mechanism design} \cite{Hurwicz, DsHamMaskin, Holmstrom, Myerson1979, Myerson1981, Myerson1982}. 
The usual approach is to design  a system in which the users make reports to the manager on the basis of their private information, the manager provides instructions to the users based on these reports, and the users then take actions that maximize their own welfare.  
A version of the \emph{revelation principle} \cite{Myerson1982} implies that such systems can always be designed so that the users find it in their own self-interest to report truthfully and act obediently. 
We merge such an approach with the innovation introduced by \cite{ParkMihaela_JSAC}, and applied to situations of medium access control \cite{ParkMihaela_EURASIP, ParkMihaela_Gamenets}, and power control  \cite{XiaoMihaela_JSTSP}, by allowing for \emph{intervention} by the manager.\footnote{A packet-dropping scheme that follows the same philosophy as intervention was proposed for flow control games in \cite{GaiKrishnamachari_Infocom}.} 
That is, we allow the manager, in addition to designing a system of reports and instructions, to deploy an \emph{intervention device} that intervenes after the users take actions. 
The action of this intervention device depends on the reports and the actions of the users, and it follows an \emph{intervention rule} designed by the manager. 
The intervention device adds to the manager's ability to provide incentives for the users to report truthfully and obey instructions by threatening \emph{punishments} if users lie and/or disobey. 

In this paper we explore the manager's problem in a class of abstract environments that exhibit some features common to many resource sharing situations in communication networks, including power control \cite{HuangBerry_JSAC06, XiaoMihaela_JSTSP}, medium access control (MAC),  
\cite{ParkMihaela_EURASIP, YiMihaela_TCOM}, and flow control \cite{GaiKrishnamachari_Infocom, YiMihaela_TCOM, BharathKumarJaffe, DouligerisMazumdar, ZhangDouligeris_TCOM}. 
We will characterize a \emph{coordination mechanism}, i.e., a system of reports, recommendation and intervention, that is optimal (from the point of view of the manager) among all mechanisms. 
We provide conditions on the environment under which it is \emph{possible} for the manager to achieve its benchmark optimum -- the outcome it could achieve if users were compliant -- and conditions under which it is \emph{impossible} for the manager to achieve its benchmark optimum.
Although we can characterize the optimal mechanism, other mechanisms are also of interest, for several reasons.
The optimal mechanism may be very difficult to compute, and hence to execute.
It is therefore of some interest to consider mechanisms that are sub-optimal but easy to compute, and we provide a simple algorithm that converges to such a mechanism. 
Moreover, in some situations, it may not be possible for the users to communicate with the manager, so it is natural to consider intervention schemes that do not require the users to make reports. 
Finally, we apply these results in the context of flow control.
Computations show, among other things, that the considered schemes can considerably increase the efficiency of the system. 

There is by now a substantial communication engineering literature that addresses the problem of providing incentives for strategic users to obey a particular resource allocation scheme.
Some of this literature adopts pricing schemes that charge users for their resource usage. 
Pricing schemes can be divided into two categories: \emph{pricing for strategic users} \cite{WangComaniciu_WPC, YangKimZhangChiangTan_INFOCOM11, BasarSrikant02, ShenBasar07} and \emph{pricing for distributed algorithms} \cite{SchmidtBerry09, HuangBerry_JSAC06, AlpcanBasar_CDMA}. 
The former is used for scenarios where the users are self-interested and strategic, as in our scenario.
Such users are required to pay \emph{real} money for their resource usage.
If the manager knows how a payment affects the utility of a user, it can give the incentives to the user to adopt a particular resource allocation scheme by setting the right prices.
Such pricing schemes may achieve the goal of optimal levels of resource usage, but suffer from the following drawbacks: (1) the users are forced, ``by contract'', to pay depending on their resource usage and on the state of the system\footnote{Current communication networks use different business models.}; (2) the manager has to know the users' monetary valuation for the service; (3) a secure infrastructure to collect the money is needed.
Pricing for distributed algorithms is used for scenarios where the users are compliant and game theory is used as a tool to obtain an efficient distributed algorithm.\footnote{This is not the scenario considered in this paper, but we want to complete the discussion on pricing schemes to remark that they might be applied to two different scenarios and to avoid misunderstandings.}
In this case the users accept passively the utilities imposed by the manager, that incorporate a term that represents a cost, even though the payments do not actually need to be carried out.
The distributed algorithm is obtained forcing the users to act as selfish agents that maximize such utilities, using for example a best response dynamic.
Game theory allows to foresee the outcome of this interaction, and the manager has to design the users' utilities to obtain a desired outcome.

A different literature, including \cite{ParkMihaela_JSAC, ParkMihaela_EURASIP, ParkMihaela_Gamenets, XiaoMihaela_JSTSP}, adopts the intervention schemes considered here. 
Intervention differs from pricing\footnote{Since in this paper we consider self-interested and strategic users, we implicitly refer to the first category of pricing schemes.} in that it operates \emph{inside} the system while pricing operates \emph{outside} the system: both schemes provide the manager with a tool to alter the utility of users, but intervention affects resource usage -- and hence utility -- directly, while pricing affects utility indirectly, through payments.
Thus, intervention is more robust than pricing: users cannot evade intervention but they might be able to evade monetary charges, moreover, the manager does not need to know the users' monetary valuation for the service in intervention schemes.

So far, both intervention and pricing schemes have mainly been applied in communication engineering games with complete information, i.e., assuming that the manager knows the relevant information held by the users.
There are few works that address the problem of extracting the relevant information from the users.
Such works (e.g., \cite{HuangBerry_06_Auction, AlpcanBocheNaik, PavanJenTara}) apply the ideas of mechanism design for auctions, creating schemes that ask the users to reveal their monetary valuation for the service and, depending on it, to pay for their resource usage.
These schemes suffer from the same defects as the previously cited pricing schemes: the users are forced to pay depending on their resource usage and a secure infrastructure to collect the money is needed.
Table \ref{tab:1} summarizes the main differences between the above described incentive schemes used in communication engineering literature and our approach. 

The remainder of this paper is organized as follows.
In Section \ref{sec:frm}, we introduce the coordination mechanism model, using Myerson's framework \cite{Myerson1982} as the reference. 
In Section \ref{sec:dm}, we study the properties of the optimal mechanism.
In Section \ref{sec:sdm}, we consider two suboptimal mechanisms which, under some assumptions, are easier to compute with respect to the optimal mechanism.
In Section \ref{sec:fcg}, we analyze and we show the results for the flow control game, both in the complete and incomplete information frameworks, and with and without intervention. 
Section \ref{sec:conc} concludes with some remarks.

\begin{table}
\begin{center}
\begin{tabular}{|m{3.5cm}|m{4cm}|m{3.1cm}|m{3cm}|}
\hline
 & \textbf{Knowledge of users' monetary valuations for the service} & \textbf{Users' behaviors in \newline reporting information} & \textbf{Users' behaviors in \newline taking actions} \\
\hline
\textbf{This work} (intervention + mechanism design) & not needed & Truthful communication \newline enforced by intervention & Actions enforced by \newline intervention \\
\hline
\textbf{Intervention} & not needed & Compliant users & Actions enforced by \newline intervention \\
\hline
\textbf{Pricing} for strategic users & needed & Compliant users & Actions enforced by \newline payments $\backslash$ contract \\ 
\hline
\textbf{Pricing} for distributed \newline algorithms & not needed & Compliant users & Compliant users \\ 
\hline
\textbf{Conventional mechanism design} (e.g., auctions) & This is the information the users are asked to report & Truthful communication \newline enforced by the scheme & Actions enforced by \newline payments $\backslash$ contract \\
\hline
\end{tabular}
\end{center}
\caption{Comparison of different incentive schemes exploited in communication engineering literature.}
\label{tab:1}
\end{table}

\section{A generalized coordination mechanism for private information problems}\label{sec:frm}

We consider a manager that wants to design a system whose resources will be used by $n$ users, $\mathcal{N} = \left\lbrace 1, 2, ..., n \right\rbrace$ denoting the set of users.
Each user might have private information that the manager cannot observe and might take an action that the manager cannot directly control.
We denote by $T_i = \left\lbrace \tau_{i, 1}, \tau_{i, 2}, ..., \tau_{i, m_i} \right\rbrace \subset \mathbb{R} $, $m_i \in \mathbb{N}$, 
%$\tau_{i, l} \in \mathbb{R}$, $l= 1,...,m_i$, 
the finite set of user $i$'s private information, in which the elements are labeled in increasing order, i.e., $\tau_{i, 1} < \tau_{i, 2} <, ..., < \tau_{i, m_i}$.
We denote by $D_i = \left[ d_i^{min}, \; d_i^{max} \right] \subset \mathbb{R}$ the set of user $i$'s possible actions. 
We refer to $t_i \in T_i$ and to $d_i \in D_i$ as the type and the action of user $i$.
As an example, each action $d_i$ may represent user $i$'s level of resource usage, while each type $t_i$ may represent $i$'s personal valuation for the resource.
We denote by $D = \times_{j \in \mathcal{N}} D_j $ and $T = \times_{j \in \mathcal{N}} T_j $ the set of joint action profiles and the set of joint type profiles, i.e., all the possible combinations of users' actions and users' types; and by $D_{-i} = \times_{j \in \mathcal{N} \setminus \left\lbrace i\right\rbrace } D_j $ and $T_{-i} = \times_{j \in \mathcal{N} \setminus \left\lbrace i\right\rbrace } T_j $ all the possible combinations of users' actions and users' types except for user $i$.
Thus, the symbols $t \in T$, $d \in D$, $t_{-i} \in T_{-i}$ and $d_{-i} \in D_{-i}$ represent vectors.

We assume that the manager can instruct a device, which we refer to as the \emph{intervention device}, that will interact with the users in the system.
The aim of the manager is to design the intervention device so that the outcome of the system maximizes the manager's objective. 
%As an example, manager objective might be the revenue or a social welfare function.
The intervention device has three features: 1) it can communicate with users; 2) it can monitor users' actions; 3) it can take an action of its own, which we interpret following \cite{ParkMihaela_JSAC} as an \emph{intervention}.
We define the \emph{intervention rule} $f: D \rightarrow D_0 = \left[ d_0^{min}, \; d_0^{max} \right]$ as a function that maps an action profile to an action of the intervention device, and we denote by $\mathcal{F}$ the finite set of intervention rules that the intervention device can implement.
For the moment (the role of the intervention device for incomplete information scenarios will be clear in Section \ref{sec:cm}), we assume that the intervention device takes an action following the randomized intervention rule $\pi$ designed by the manager, such that $\pi(f) \geq 0 \, , \, \forall \, f \in \mathcal{F}$, and $\sum_{f\in \mathcal{F}} \pi(f) = 1$.
The randomized intervention rule is communicated to all users before they select their actions.
After observing users' actions $d$, the intervention device picks an intervention rule $f$ following the probability distribution $\pi$ and intervenes with an action $f(d)$.
We refer to the couple $\left( D_0 , \mathcal{F} \right) $ as the intervention capability. 
Finally, we denote by $U_0: \mathcal{F} \times D \times \mathcal{T} \rightarrow \mathbb{R}$ the manager's utility function, 
and by $U_i: \mathcal{F} \times D \times \mathcal{T} \rightarrow \mathbb{R}$ user $i$'s utility function, where $\mathcal{T} = \times_{i \in \mathcal{N}} \mathcal{T}_i$, $\mathcal{T}_i = \left[ \tau_{i, 1}, \; \tau_{i, m_i} \right]$.\footnote{We require the manager's utility to be defined over the continuous interval $\mathcal{T} \subset \mathbb{R}^n$, that includes the finite type set $T$, because the property \textbf{A3} needs a set in which the differentiation operation is defined.
However, the results in the rest of the paper are obtained under the condition that each user $i$'s type belongs to the finite set $T_i$.}
%and $T_i \subseteq \mathcal{T}_i$, $\forall i \in \mathcal{N}$.
%With this notation we want to indicate that the utility functions are well defined also if the third argument, representing the type profile, belongs to a continuous interval $\mathcal{T}$ of $\mathbb{R}^n$, which includes
%, but may be different from, 
%the set of type profiles set $T$.  

\subsection{Assumptions on utilities}\label{sec:ut}

We assume that the manager's utility satisfies the following assumptions, $\forall d \in D$ and $\forall t \in \mathcal{T}$,
\begin{itemize}
\item[\textbf{A1}:] There exists $d_0^* \in D_0$ such that $U_0 (d_0^*, d, t) > U_0 (d_0, d, t)$, $\forall \, d_0 \in D_0$, $d_0 \neq d_0^*$
\item[\textbf{A2}:] $d^*(t) = \argmax_d U_0(d_0^*, d, t)$ is unique
\item[\textbf{A3}:] $d_i^*(t)$ is differentiable with respect to $t_i$ and $\frac{\partial d_i^*(t)}{\partial t_i} > 0$
\end{itemize}

Assumption \textbf{A1} states that $d_0^*$ is the most preferred action of the manager, regardless of users' actions and type profile. 
In games where the intervention device drives users' actions by threatening punishments, the intervention can be interpreted as the level of punishment and $d_0^*$ as the absence of intervention.

By assumption \textbf{A2}, for every type profile $t \in \mathcal{T}$ and for every user $i \in \mathcal{N}$, the users' joint action profile that maximizes the intervention device's utility is unique, and by assumption \textbf{A3},
%(e.g., $U_0(d_0^*, d, t)$ concave in $d$ respects this assumption) 
each component in $d^*(t)$ is continuous and increasing in the type of that user.
If actions represent the level of resource usage and types represent resource valuations, assumption \textbf{A3} asserts that the higher $i$'s valuation the higher should be $i$'s level of resource usage.

For each type profile $t \in \mathcal{T}$, we define the game 
\begin{equation}
\Gamma^0_t = \left( \mathcal{N}, D, \left\lbrace U_i(d_0^*, \cdot, t) \right\rbrace_{i=1}^n  \right)
\label{eq:gamma_0_t}
\end{equation}
$\Gamma^0_t$ is the complete information game (i.e., users know everything about the structure of the game, in particular, they know the types of the other users) that models the interaction between strategic users having types $t$ when the intervention device adopts the action $d_0^*$ independently of users' actions.
It can be thought as the complete information game that models users' interaction in the absence of an intervention device.

We denote by $d^{NE^0}(t) = \left( d_1^{NE^0}(t), \dots, d_n^{NE^0}(t) \right)$ a Nash Equilibrium ($NE$) of the game $\Gamma^0_t$, which is an action profile so that each user obtains its maximum utility given the actions of the other users, i.e., 
\begin{equation}
U_i \left( d_0^*, d^{NE^0}(t), t \right) \geq U_i \left( d_0^*, d_i, d_{-i}^{NE^0}(t), t \right) \;\; , \;\; \forall \, d_i \in D_i \; , \; \forall \, i \in \mathcal{N} 
\end{equation} 
Notice that we have a different game $\Gamma^0_t$, and therefore a different $NE$ action profile, for each possible type profile $t \in T$. 
For this reason $d^{NE^0}(t)$ is represented as a function of $t$.

We assume that users' utilities $U_i (d_0^*, d, t)$ are twice differentiable with respect to $d$ and, $\forall \, d \in D$, $\forall \, t \in T$, $\forall \, i, j \in \mathcal{N}$, $i \neq j$,
\begin{itemize}
\item[\textbf{A4}:] $U_i (d_0^*, d, t)$ is quasi-concave in $d_i$ and there exists a unique best response function $d^{BR}_i(d_{-i}, t) = \argmax_{d_i} U_i (d_0^*, d, t)$
\item[\textbf{A5}:] $\frac{\partial^2 U_i (d_0^*, d, t)}{\partial d_i \partial d_j} \leq 0$
\item[\textbf{A6}:] There exists $d^{NE^0}(t)$ such that $d^{NE^0}(t) \geq d^*(t)$ \footnote{Throughout the paper, inequalities between vectors are intended component-wise.} and $d_k^{NE^0}(\tau_k, t_{-k}) > d_k^*(\tau_k, t_{-k})$ for some users $k \in \mathcal{N}$ and type $\tau_k \in T_k$
%and $\exists (k, \tau_k) \in \mathcal{N} \times T_k$ such that $d_k^{NE^0}(\tau_k, t_{-i}) > d_k^*(\tau_k, t_{-i})$
\end{itemize}

Assumption \textbf{A4} states that $\Gamma^0_t$ is a quasi-concave game and the best response function $d^{BR}_i(d_{-i}, t)$ that maximizes $U_i (d_0^*, d, t)$ is unique.
Hence, either $i$'s utility is monotonic with respect to $d_i$, or it increases with $d_i$ until it reaches 
a maximum for $d^{BR}_i(d_{-i}, t)$, and decreases for higher values.
As a consequence, a NE $d^{NE^0}(t)$ of $\Gamma^0_t$ exists. 
In fact, the best response function $d^{BR}(d, t) = \left( d^{BR}_1(d_{-i}, t), \dots, d^{BR}_n(d_{-i}, t) \right) $ 
is a continuous function from the convex and compact set $D$ to $D$ itself, therefore Brouwer's fixed point theorem 
assures that a fixed point exists.

Assumption \textbf{A5} asserts that $\Gamma^0_t$ is a submodular game and it ensures that $d^{BR}_i(d_{-i}, t)$ 
is a non increasing function of $d_j$.
Interpreting $d_i$ as $i$'s level of resource usage, this situation reflects resource allocation games where it is in the interest of a user not to increase its resource 
usage if the total level of use of the other users increases, in order to avoid an excessive use of the resource.
Nevertheless, assumption \textbf{A6} says that strategic users use the resources more heavily compared to the optimal 
(from the manager's point of view) usage level.

The class of games satisfying assumptions \textbf{A4}-\textbf{A6} includes the linearly coupled games \cite{YiMihaela_TCOM} 
and many resource allocation games in communication networks, such as the MAC 
\cite{ParkMihaela_EURASIP, YiMihaela_TCOM}, power control \cite{HuangBerry_JSAC06, XiaoMihaela_JSTSP} 
and flow control \cite{GaiKrishnamachari_Infocom, YiMihaela_TCOM, BharathKumarJaffe, DouligerisMazumdar, ZhangDouligeris_TCOM} games.
Moreover, if the manager's utility is increasing in the users' utilities (e.g., sum-utilities or geometric mean) 
and the intervention represents a punishment, also assumptions \textbf{A1}-\textbf{A3} are satisfied in these games and the absence of intervention represents the intervention device's preferred action $d_0^*$.

\subsection{Actions enforcement for the complete information game}\label{sec:ci}

We first introduce the framework to design incentives to enforce users' actions in the complete information scenario, though the main focus of this paper is the design of a system for an incomplete information setting, dealing both with information revelation and action enforcement.
% in presence of strategic users having private information
The notations and concepts introduced in the following will become useful later, when we study the incomplete information scenario.
In fact, some properties of the incomplete information game (i.e., the game where users do not know the types of the other users) are linked to the properties of the complete information game defined in this Subsection.

Given a randomized intervention rule $\pi$, we define the complete information game 
\begin{equation}
\Gamma_t = \left( \mathcal{N}, D, \left\lbrace \overline{U}_i(\cdot, t) \right\rbrace_{i=1}^n \right)
\label{eq:gamma_t}
\end{equation}
that models the interaction between strategic users having types $t$.
%The strategy profile $\delta \in \Delta $ is a vector of functions $\delta_i: T_i \rightarrow D_i $ that assign to each user user $i$ an action depending on $i$'s type, $i \in \mathcal{N}$.
The utility functions $\overline{U}_i(\cdot, t)$ are the expectations, over the randomized intervention rule, of the original utilities:
\begin{equation}
\overline{U}_i \left( d, t \right) =  \mathbb{E}_{f} \left[ U_i \left( f, d, t \right) \right] =
\sum_{f \in \mathcal{F}} \pi \left( f \mid t \right) U_i \left( f, d, t \right) 
\label{eq:U_i_over}
\end{equation}
where $\pi \left( f \mid t \right)$ denotes the probability that the intervention device adopts the intervention rule $f \in \mathcal{F}$ given that the type profile is $t$, and $\mathbb{E}_x \left[ \cdot \right]$ is the expectation operator with respect to the random variable $x$.\footnote{There is some abuse of notation in using the same symbol to indicate a random variable and a particular realization, but this will not lead to confusion.}

Analogously, we denote by $\overline{U}_0$ the manager's expected utility
\begin{equation}
\overline{U}_0 \left( d, t \right) =  \mathbb{E}_{f} \left[ U_0 \left( f, d, t \right) \right] =
\sum_{f \in \mathcal{F}} \pi \left( f \mid t \right) U_0 \left( f, d, t \right) 
\label{eq:U_0_over}
\end{equation}
According to assumptions \textbf{A1}-\textbf{A2}, the manager's expected utility is maximized when users adopt action profile $d^*(t)$ and the intervention device adopts action $d_0^*$.
%, i.e., for every intervention rule $f$ that is adopted with positive probability, it must be $f \left( d^*(t) \right) = d_0^*$.
% $\pi \left( f \mid t \right) > 0$ $f \left( d^*(t) \right) = d_0*$, for every intervention rule $f$ that is adopted with positive probability, i.e., such that $\pi \left( f \mid t \right) > 0$. 
However, in a strategic scenario the users adopt the actions that maximize their own utilities, and the possible outcomes are represented by the $NEs$.
The $NEs$ of the game $\Gamma_t$ depend on the randomized intervention rule selected by the manager because it affects the utilities of the users.
Thus, the manager has to design the randomized intervention rule so that there exists a $NE$ of the game $\Gamma_t$
that gives it the highest utility among what is achievable with all possible $NEs$.
%If the set of users using the system changes over the time, then the intervention device has to design a randomized intervention rule for each possible type profile $t$.

\begin{defn}
A randomized intervention rule $\pi$ is said to sustain an action profile $d \in D$ in $\Gamma_t$ if $d$ is a $NE$ of the game $\Gamma_t$, i.e., if 
%$\pi \left( \hat{f} \mid t \right) = 1$, 
\begin{align}
\overline{U}_i \left( d, t \right) \geq \overline{U}_i \left( \hat{d}_i, d_{-i}, t \right) \;\; , \;\; \forall \, i \in \mathcal{N} \; , \; \forall \, \hat{d}_i \in D_i
%\sum_{f \in \mathcal{F}} \pi \left( f \mid t \right) U_i \left( f, d, t \right) \geq \sum_{f \in \mathcal{F}} \pi \left( f \mid t \right) U_i \left( f, \hat{d}_i, d_{-i}, t \right) \;\; , \;\; \forall \, i \in \mathcal{N} \; , \; \forall \, \hat{d}_i \in D_i
\label{eq:V_cos}
\end{align}
If such $\pi$ exists, we say that $d$ is sustainable.

A randomized intervention rule $\pi$ sustains an action profile $d \in D$ in $\Gamma_t$ without intervention if $\pi$ sustains $d$ and $f(d) = d_0^*$ for every intervention rule $f$ such that $\pi \left( f \mid t \right) > 0$.\footnote{These definition can be easily extended for \emph{pure} intervention rule: $\hat{f}$ sustains $d$ in $\Gamma_t$ (without intervention) if $\pi$ sustains $d$ in $\Gamma_t$ (without intervention), where $\pi \left( f \mid t \right) = 1$ if $f=\hat{f}$, $0$ otherwise.}
If such $\pi$ exists, we say that $d$ is sustainable without intervention.

%A \emph{pure} intervention rule $\hat{f} \in \mathcal{F}$ sustains an action profile $d \in D$ in $\Gamma_t$ (without intervention) if $\hat{\pi}$ sustains $d$ in $\Gamma_t$ (without intervention), where $\hat{\pi} \left( f \mid t \right) = 1$ if $f = \hat{f}$, $0$ otherwise.
\end{defn}

Interpreting $d_0^*$ as the absence of intervention, the expression \emph{sustainable without intervention} is here used to indicate that in the equilibrium the intervention action is not  executed.
We denote by $\mathcal{F}^{d,t}$ the set of all randomized intervention rules, obtainable starting from the intervention rule set $\mathcal{F}$, that sustain $d$ in $\Gamma_t$ without intervention.
The possibility of the manager to design a randomized intervention rule capable of sustaining an action profile depends on the intervention capability, namely, the action space $D_0$ of the intervention device and the class of intervention rules $\mathcal{F}$ the intervention device is able to implement.
If we expand these sets, the manager has more degrees of freedom in designing intervention rules capable of sustaining action profiles.

\begin{defn}
$\left( D_0, \mathcal{F} \right)$ is an optimal intervention capability with respect to the complete information game $\Gamma_t$ if the maximum utility that the intervention device can obtain considering all the sustainable action profiles cannot be improved by expanding $D_0$ and $\mathcal{F}$.
\label{def:opt_class}
\end{defn}

\subsection{Coordination mechanism formulation for the incomplete information game}\label{sec:cm}

In this paper we consider the scenario where each user has private information, which is synthesized in its type. 
Following Harsanyi's approach \cite{Harsanyi}, we study the incomplete information scenario assuming that each user acts based on the beliefs it has about the types of the other users.
In particular, we denote by $P_t(\cdot)$ the joint probability distribution of the type profile over the type profile set $T$.
%(in case $T$ is continuous, $P_t(\cdot)$ must be interpreted as a density function). 
We assume that each type profile has a positive probability to occur, i.e., $P_t(\tau) > 0$, $\forall \, \tau \in T$.  
We denote by $P_{t_{-i}}(\tau_{-i})$ the joint probability distribution of the type profile of all the users except for user $i$ over the set $T_{-i}$ (notice that user $i$ knows its own type, $t_i$).
We assume that, for each user $i$, $P_{t_{-i}}(\tau_{-i})$ is consistent with $P_t(\cdot)$, i.e., $P_{t_{-i}}(\tau_{-i}) = P_t(t_i, \tau_{-i} \mid t_i)$.

To reach its objective, the manager may program the intervention device to elicit information from users and to spread information into the system (notice that users' behaviors, and therefore the outcome of the system, depend on the information they have).
%However, in doing so, the manager has to take into consideration that users may strategically lie when asked to report information.
We denote by $R_i$ the set of all reports that user $i$ can transmit to the intervention device and by $M_i$ the set of all messages the intervention device can send to user $i$.
As usual, we denote by $r_i \in R_i$ the report sent by $i$, by $r \in R = \times_{i \in \mathcal{N}} R_i$ the report profile, by $m_i \in M_i$ the message sent to user $i$ and by $m \in M = \times_{i \in \mathcal{N}} M_i$ the message profile.
The messages sent and the randomized intervention rule adopted by the intervention device may depend on the reports sent by users. 
Hence, given the report profile $r$, we denote by $m^S(r) = (m^S_1(r), \dots, m^S_n(r))$, $m^S_i: R \rightarrow M_i$, the messages sent by the intervention device and by $\pi \left( f \mid r \right)$ the probability that the intervention rule $f \in \mathcal{F}$ is adopted.
Following Myerson's terminology \cite{Myerson1982}, we refer to $\left( R, M, m^S, \pi \right)$ as the \emph{coordination mechanism} implemented by the intervention device. 

The manager has to design the coordination mechanism to drive the outcome of the system towards its objective.
In doing so, it has to consider that users might both send reports and adopt actions strategically, i.e., both information revelation and action enforcement issues must be addressed at the same time.
Once the coordination mechanism is established, the interaction between users can be modeled as a Bayesian game  
\begin{equation}
\Gamma = \left( \mathcal{N}, \Phi, \Delta, T, P_t, \left\lbrace \overline{U}_i(\cdot, \cdot, t) \right\rbrace_{i=1}^n  \right)
\label{eq:gamma}
\end{equation}
In this context, a strategic user $i$ selects its report $r_i \in R_i$ and its action $d_i \in D_i$ in order to maximize its expected utility given the information and the beliefs it has.
Precisely, a strategy for user $i$ consists of a couple of functions $\left( \phi_i, \delta_i \right)$.
$\phi_i : T_i \rightarrow R_i$ represents the report of user $i$ which may depend on its type.
$\delta_i: M_i \times T_i \rightarrow D_i$ represents the action of user $i$ which may depend on its type and on the message received; in fact the received message can carry information about the types of the other users, that can be exploited by $i$ to select the most appropriate action.
We denote by $\phi = \left\lbrace \phi_i \right\rbrace_{i \in \mathcal{N}} \in \Phi$ the reporting strategy profile and by $\delta = \left\lbrace \delta_i \right\rbrace_{i \in \mathcal{N}} \in \Delta$ the action strategy profile.

\begin{figure}[p]%[htp]
     \centering
          \includegraphics[width=\figw]{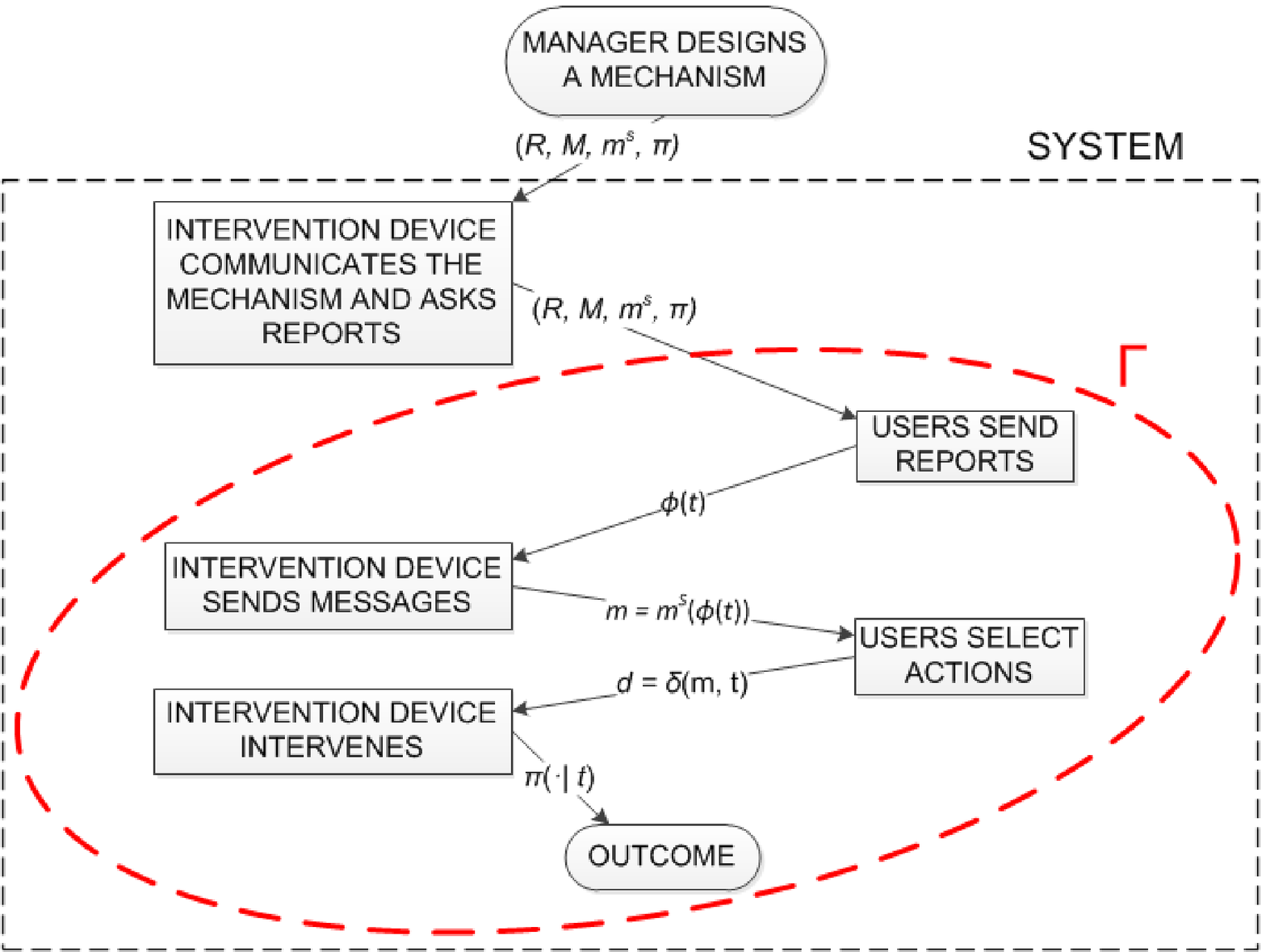}
\caption{Interaction between the users and the intervention device}
\label{fig:md}
\end{figure}

Fig. \ref{fig:md} represents the different stages of the interaction between the users and the intervention device, which are summarized in the following.
\begin{itemize} 
\item[\textbf{Stage 1}:] the intervention device announces the coordination 
mechanism $\left( R, M, m^S, \pi \right)$ \footnote{We remark the importance of communicating the 
mechanism and committing to it. If the intervention device could deviate from the mechanism and select an action to maximize the manager's utility, then, since \textbf{A1} is satisfied, the intervention device would adopt $d_0^*$ independently of users' actions. The users, foreseeing this behavior, would ignore the threat of the intervention device and would play as if the intervention device were not present in the system. Conversely, forcing the intervention device to follow the mechanism and communicating it to users, allows the manager to design credible threats and obtain better outcomes.}
\item[\textbf{Stage 2}:] each user $i$ sends a report $\phi_i(t)$ to the intervention device 
\item[\textbf{Stage 3}:] the intervention device sends a message $m_i = m^S_i \left( \phi_i(t)\right) $ to each user $i$
\item[\textbf{Stage 4}:] each user $i$ takes an action $d_i = \delta_i \left( m_i, t_i \right) $
\item[\textbf{Stage 5}:] the intervention device monitors the users' action profile $d$, picks an intervention rule $f$ following the distribution $\pi \left( \cdot \mid \phi(t) \right)$, and adopts the action $f(d)$
\end{itemize}

%%[p] mette le figure alla fine dell'articolo
%\begin{figure}[p]%[htp]
%     \centering
%          \includegraphics[width=\figw]{Mechanism_noFig2.eps}
%\caption{Interaction between the users and the intervention device}
%\label{fig:md}
%\end{figure}

The utility $\overline{U}_i$ of each user is the expectation over the randomized intervention rule of the original utilities, therefore, given the strategy profiles $\left( \phi, \delta \right)$,
\begin{equation}
\overline{U}_i \left( \phi, \delta, t \right) =  \mathbb{E}_{f} \left[  U_i \left( f, \delta(m^S(\phi(t)),t), t \right) \right] =
\sum_{f \in \mathcal{F}} \pi \left( f \mid \phi(t) \right) U_i \left( f, \delta(m^S(\phi(t)),t), t \right)
\label{eq:U_i_over_bne}
\end{equation}

In a Bayesian game a user selects its strategy in order to maximize the expectation of its utility with respect 
to the initial beliefs about the types of the other players. 
The expected utility of a user $i$ having type $t_i$ is
\begin{equation}
V_i \left( \phi, \delta, t_i \right) = \mathbb{E}_{t_{-i}\mid t_i} \left[  \overline{U}_i \left( \phi, \delta, t \right) \right] =
\sum_{t_{-i} \in T_{-i}} \sum_{f \in \mathcal{F}} P_t(t \mid t_i) \pi \left( f \mid \phi(t) \right) U_i \left( f, \delta(m^S(\phi(t)),t), t \right) 
\label{eq:V}
\end{equation}

The strategy profiles $\left( \phi, \delta \right)$ is a Bayesian Nash Equilibrium ($BNE$) of the game if, for each user $i \in \mathcal{N}$, for every type $t_i \in T_i$ and for every alternative strategy $\left( \tilde{\phi}_i, \tilde{\delta}_i \right)$ for $i$,
\begin{align}
V_i \left( \phi, \delta, t_i \right) \geq V_i \left( \tilde{\phi}_i, \phi_{-i}, \tilde{\delta}_i, \delta_{-i}, t_i \right)
\label{eq:V_cos}
\end{align}

Finally, the aim of the manager is to design an optimal coordination mechanism $\left( R, M, m^S, \pi \right)$, such that there is a $BNE$ $\left( \phi, \delta \right)$ that gives the manager the highest possible expected utility
\begin{equation}
V_0 \left( \phi, \delta \right) = \mathbb{E}_t \left[  \mathbb{E}_{f} \left[ U_0 \left( f, \delta(m^S(\phi(t)),t), t \right) \right] \right] =
\sum_{t \in T} \sum_{f \in \mathcal{F}} P_t(t) \pi \left( f \mid \phi(t) \right) U_0 \left( f, \delta(m^S(\phi(t)),t), t \right) 
\label{eq:V0}
\end{equation}

This formulation is rather abstract, so it may be worth to use a simple illustrative example to remark our goal.
Assume the manager has to assign a resource to 2 users. 
From a social point of view, the best choice might be to assign the resource to the user having the higher valuation for that resource.
Using a conventional MD scheme, the manager might implement a \emph{Vickrey auction} to obtain the users' valuations and to select the user with the higher valuation. 
However, if a user could avoid the payment such method would fail its objective because that user could bid more than what it is really willing to pay. 
Moreover, nothing would prevent the user that has lost the auction from trying to access the resource. 
That is, conventional mechanism design relies on other systems (e.g., a reliable infrastructure to collect money and punishments for the users that do not respect the agreements) to be effective.
Here we want to design a scheme that does not rely on external systems. 
As an example, the intervention device might be a device that asks the users to report their valuations and, based on that, proposes how to share the resource.
If the users do not respect such sharing the intervention device might jam their communication. 
The mechanism used by the intervention device to propose the resource sharing and to jam users' communication must be designed to provide the incentive for both the users to report their true valuations and to accept the proposed resource sharing.

\section{Optimal incentive compatible direct mechanisms}\label{sec:dm}

The design of an optimal coordination mechanism seems to be intractable since there are no constraints on the sets $M_i$ and $R_i$.
Fortunately, the revelation principle \cite{Myerson1982} allows us to restrict the attention to the class of \emph{incentive compatible direct mechanisms}, among which the optimal mechanism is also optimal in the class of all coordination mechanisms.
In a direct mechanism users report their types to the intervention device, and the intervention device sends them  a suggested action profile, i.e., $R_i = T_i$ and $M_i = D_i$,  $\forall i \, \in \mathcal{N}$.
We denote by $d^S(r) = (d^S_1(r), \dots, d^S_n(r))$, $d^S_i: T \rightarrow D_i$, the suggested action profile given the reported type profile $t$.
We say that user $i$ is honest and obedient if it reports its real type and adopts the suggested action, i.e., if $\phi_i (t_i) = t_i$ and $\delta_i (d_i, t_i) = d_i$, for every type $t_i \in T_i$ and suggested action $d_i \in D_i$.
Finally, a direct mechanism is incentive compatible if the honest and obedient strategy profile is a $BNE$, i.e., if it provides incentives for users to behave honestly and obediently.

The Optimal Incentive Compatible Direct Mechanism (\textbf{OICDM}) can be computed solving  
\begin{align}
%& \argmax_{d^S, \pi} V_0 \left( \phi^*, \delta^* \right) = 
\textbf{OICDM} \;\;\;\;\; &\argmax_{d^S, \pi} \sum_{t \in T} \sum_{f \in \mathcal{F}} P_t(t) \pi \left( f \mid t \right) U_0 \left( f, d^S(t), t \right) \nonumber\\
& \mbox{subject to:} \nonumber\\
&\pi \left( f \mid t \right) \geq 0 \;\; , \;\; \sum_{x \in \mathcal{F}} \pi \left( x \mid t \right) = 1 \;\; , \;\; \forall f \in \mathcal{F} \; , \; \forall t \in T \nonumber\\
&\sum_{t_{-i} \in T_{-i}} \sum_{f \in \mathcal{F}} P_t(t \mid \tau_i) \pi \left( f \mid t \right) U_i \left( f, d^S(t), t \right) \geq \nonumber\\
& \;\;\;\;\;\;\;\;\;\;\;\;\;\; \geq \sum_{t_{-i} \in T_{-i}} \sum_{f \in \mathcal{F}} P_t(t \mid \tau_i) \pi \left( f \mid t_{-i}, \hat{\tau}_i \right) U_i \left( f, d_{-i}^S(\hat{\tau}_i, t_{-i}), \hat{\delta}_i(d^S_i(t_{-i}, \hat{\tau}_i)), t \right) \nonumber\\
& \forall \, i \in \left\lbrace 1, ..., n \right\rbrace, \;\; \forall \, \tau_i \in T_i, \;\; \forall \, \hat{\tau}_i \in T_i, \;\; \forall \, \hat{\delta}_i : D_i \rightarrow D_i \nonumber
\end{align}

The second set of constraints of \textbf{OICDM} represents the incentive compatible condition.
It asserts that when $i$'s type is $\tau_i$, $i$ does at least as well by being honest and obedient as by reporting $\hat{\tau}_i$ and then adopting $\hat{\delta}_i(d^S_i(t_{-i}, \hat{\tau}_i))$ when told to adopt $d^S_i(t_{-i}, \hat{\tau}_i)$, assuming that the other users are honest and obedient.
If users were compliant to the manager's instructions, the mechanism could be thought as a way to retrieve the relevant information, compute the optimal policy and recommend actions to users. 
In this scenario the optimal mechanism could be computed solving \textbf{OICDM} without the second set of constraints.
The design of a system that is robust against self-interested strategic users translates mathematically in additional constraints to satisfy, which represent the incentives given to users to follow the instructions.
For this reason, the maximum utility the manager can obtain with self-interested strategic users is never higher than the maximum utility it can achieve with compliant users.
We denote by $V_0^{ME}$ the maximum expected utility that the manager can obtain when users are compliant, i.e., 
%For games satisfying \textbf{A1}-\textbf{A2}, we can write:
\begin{align}
V_0^{ME} = \sum_{t \in T} P_t(t) U_0(d_0^*, d^*(t), t)  
\label{eq:v0_ME}
\end{align}
We say that a direct mechanism is a maximum efficiency incentive compatible direct mechanism if it is a solution of \textbf{OICDM} and the expected utility that the manager can achieve is equal to the maximum efficiency utility.

Finally, we define the concept of optimal intervention capability also for the incomplete information game $\Gamma$.
\begin{defn}
$\left( D_0,  \mathcal{F} \right)$ is an optimal intervention capability with respect to the incomplete information game $\Gamma$ if the solution of \textbf{OICDM} cannot be improved expanding $D_0$ and $\mathcal{F}$.
%In this case, we say that the solution of Eq. (\ref{eq:in_comp}) is an optimal incentive compatible solution.
\label{def:opt_class2}
\end{defn}

%In the following Subsection we discuss the existence and characterize the computation of a maximum 
%efficiency incentive compatible direct mechanism.
%Such a mechanism might not exists, therefore it is important to derive a solution 
%of \textbf{OICDM} in the general case.
%In Subsection \ref{sec:po} we will see that, with some 
%additional assumptions, \textbf{OICDM} can be simplified.
%Finally, in Section \ref{sec:sdm}, we will propose two suboptimal mechanisms that, while unable  
%to reach the same performance as the solution of 
%\textbf{OICDM}, are much easier to compute.

\subsection{Properties of a maximum efficiency incentive compatible direct mechanism}\label{sec:prop}

In this Subsection we address the problem of the existence and the computation of a maximum efficiency incentive compatible direct mechanism.

The first result we derive asserts that a maximum efficiency incentive compatible direct mechanism exists if and only if, for every type profile $t$, the optimal action profile $d^*(t)$ is sustainable in the game with complete information $\Gamma_t$, and users have incentives to reveal their real type given that they will adopt $d^*(t)$ and the intervention device does not intervene.
If this is the case, we are also able to characterize all maximum efficiency incentive compatible direct mechanisms.

\begin{proposition}
$\left( T, D, d^S, \pi \right)$ is a maximum efficiency incentive compatible direct mechanism
if and only if, $\forall t \in T$,
\begin{itemize}
\item[\textbf{1}:] the optimal action profile $d^*(t)$ of the game $\Gamma_t$ is sustainable without intervention in $\Gamma_t$;
\item[\textbf{2}:] each user has incentives to report its real type, when other users do it and everybody is adopting the optimal action profile $d^*(t)$ and the intervention device never intervenes, i.e,
\begin{align}
&\sum_{t_{-i} \in T_{-i}} P_t(t \mid \tau_i) U_i \left( d_0^*, d^*(t), t \right) \geq \sum_{t_{-i} \in T_{-i}}  P_t(t \mid \tau_i) U_i \left( d_0^*, d^*(\hat{\tau}_i, t_{-i}), t \right) \nonumber\\
& \forall \, i \in \mathcal{N}, \;\; \forall \, \tau_i \in T_i, \;\; \forall \, \hat{\tau}_i \in T_i, 
\label{eq:in_comp2}
\end{align}
\item[\textbf{3}:] the suggested action profile is the optimal action profile of game $\Gamma_t$, i.e., $d^S(t) = d^*(t)$;
\item[\textbf{4}:] the randomized intervention rule sustains without intervention $d^*(t)$ in $\Gamma_t$, i.e., $\pi(\cdot \mid t) \in \mathcal{F}^{d^*(t),t}$.
\end{itemize}
\label{lemma1}
\end{proposition}
\begin{IEEEproof}
See Appendix \ref{app:contradiction}
\end{IEEEproof}
Conditions \textbf{1}-\textbf{2} are related to the structure of the game without intervention device, while conditions \textbf{3}-\textbf{4} say how to obtain a maximum efficiency direct mechanism once \textbf{1}-\textbf{2} are satisfied.

In the second result we combine condition \textbf{2} of Proposition \ref{lemma1} with assumptions \textbf{A3}-\textbf{A6} to derive a sufficient condition on users' type set structures under which a maximum efficiency incentive compatible direct mechanism does not exist.
We define the bin size $\beta_k$ of user $k$'s type set, $T_k$, as the maximum distance between two consecutive elements of $T_k$: $\beta_k = \max_{s \in \left\lbrace 1, \dots, m_k-1 \right\rbrace } \left( \tau_{k,s+1} - \tau_{k,s} \right)$.
We define the bin size $\beta$ as the maximum between the bin sizes of all users: $\beta = \max_{k \in \mathcal{N} } \beta_k$.

\begin{proposition}
There exists a threshold bin size $\zeta > 0$ so that if $\beta \leq \zeta$ then a maximum efficiency incentive compatible direct mechanism does not exist.
\label{prop:1}
\end{proposition}
\begin{IEEEproof}
Let $k \in \mathcal{N}$ and $\tau_k \in T_k$ be such that $d_k^{NE^0}(\tau_k, t_{-i}) > d_k^*(\tau_k, t_{-i})$, $\forall \, t_{-i} \in T_{-i}$.
We rewrite condition \textbf{2} of Proposition \ref{lemma1} for users $k$ and type $\tau_k$:
\begin{align}
&\sum_{t_{-k} \in T_{-k}} P_t(t \mid \tau_k) U_i \left( d_0^*, d^*(t), t \right) \geq \sum_{t_{-k} \in T_{-k}}  P_t(t \mid \tau_k) U_i \left( d_0^*, d^*(\hat{\tau}_k, t_{-k}), t \right) \;\; , \;\; \forall \, \hat{\tau}_k \in T_k, 
\label{eq:dis1}
\end{align}

We have $d_k^{BR}(d_{-k}^*, t_{-k}, \tau_k) \geq d_k^{BR}(d_{-k}^{NE^0}, t_{-k}, \tau_k) = d_k^{NE^0}(t_{-k}, \tau_k) > d_k^*(t_{-k}, \tau_k)$, where the first inequality is valid for the submodularity.

Let $\tilde{\tau}_k(t_{-k})$ be the type $\tau$ so that $d^*(\tau, t_{-k}) = d_k^{BR}(d_{-k}^*, t_{-k}, \tau_k)$ if it exists (in this case \textbf{A3} guarantees it is greater than $\tau_k$) and it is lower than $\overline{t}_k$, and $\tilde{\tau}_k(t_{-k}) = \overline{t}_k$ otherwise.
Let $\hat{\tau}_k = \min_{t_{-k}} \tilde{\tau_k}(t_{-k})$.
If $\left( \tau_k, \hat{\tau}_k \right] \bigcap T_k \neq \emptyset$ (in particular, this is true if $\beta \leq \hat{\tau}_k - \tau_k = \zeta $), $\forall \tau_m \in \left( \tau_k, \hat{\tau}_k \right] \bigcap T_k$ we obtain 
\begin{align}
U_k \left( d_0^*, d^*(t_{-k}, \tau_m), t_{-k}, \tau_k \right) > U_k \left( d_0^*, d^*(t_{-k}, \tau_k), t_{-k} \tau_k \right) \; , \; \forall \, t_{-k} \in T_{-k}
\label{eq:dis2}
\end{align}
contradicting Eq. (\ref{eq:dis1}).
\end{IEEEproof}

\textit{Interpretation:} when user $k$'s type is $\tau_k$, $k$'s resource usage that maximizes the manager's utility, $d_k^*(\tau_k, t_{-k})$, is lower than the one that maximizes $k$'s utility, $d_k^{BR}(d_{-k}^*, \tau_k, t_{-k})$, $\forall \, t_{-k} \in T_{-k}$. 
If $k$ reports a type $\tau_m$ slightly higher than $\tau_k$, then the intervention device suggests a slightly higher resource usage, allowing $k$ to obtain a higher utility. Hence, $k$ has an incentive to cheat and resources are not 
allocated as efficiently as possible. To avoid this situation, the intervention device might decrease the 
resources given to a type $\tau_m$. In this case the loss of efficiency occurs when the real type of $k$ is $\tau_m$ and it does not receive the resources it would deserve. There is no way to avoid the loss of efficiency associated to both case $t_k = \tau_k$ and case $t_k = \tau_m$, both occurring with positive probability.

It is worth noting that we consider finite type sets and a finite intervention rule set mainly to simplify the logical exposition.
However, all results might be derived also with infinite and continuous sets.\footnote{For the continuous case, probability distributions and sums must be substituted with probability density functions and integrals.}
In particular, if type sets are continuous Proposition \ref{prop:1} implies that a maximum efficiency incentive compatible direct mechanism never exists.

\subsection{Properties of optimal incentive compatible direct mechanisms}\label{sec:po}

If a maximum efficiency incentive compatible direct mechanism exists, the optimal incentive compatible direct mechanisms set coincides with the maximum efficiency incentive compatible direct mechanisms set, that is characterized in Proposition \ref{lemma1}.
However, finding an optimal incentive compatible direct mechanism in the general case, solving \textbf{OICDM}, may be computationally hard.
In this Subsection we consider some additional conditions to simplify the problem.
First we assume that the manager's utility is a function of the users' utilities.
Moreover, we suppose that the intervention capability $\left( D_0, \mathcal{F} \right)$ is such that, for each type profile $t \in T$, every action profile $d \in D$ lower than the $NE$ action profile of the game $\Gamma^0_t$ is sustainable without 
intervention in $\Gamma_t$ (i.e., $d \leq d^{NE^0}(t)$ implies $\mathcal{F}^{d,t}$ non empty).
Finally, we assume that, for each type profile $t \in T$ and for every action profile $d \in D$, 
the utility of a user $i$ adopting the lowest action $d_i^{min}$ is equal to $0$, i.e., $U_i (d_0^*, d_i^{min}, d_{-i}, t) = 0$.
Interpreting $d_i^{min}$ as no resource usage, this means that, independently of types and other users' actions, a user that does not use resources obtains no utility. 

\begin{lemma}
The utility of user $i$ is non increasing in the actions of the other users. 
\end{lemma}
\begin{IEEEproof}
\begin{align}
&U_i (d_0^*, d, t) = U_i (d_0^*, 0, d_{-i}, t) + \int_0^{d_i} \dfrac{\partial U_i (d_0^*, x, d, t)}{\partial x} \partial x = \int_0^{d_i} \dfrac{\partial U_i (d_0^*, x, d, t)}{\partial x} \partial x \nonumber\\
&\dfrac{\partial U_i (d_0^*, d, t)}{\partial d_j} = \int_0^{d_i} \dfrac{\partial^2 U_i (d_0^*, x, d, t)}{\partial x \partial d_j} \partial x \leq 0
\label{eq:dec_from_sub}
\end{align}
where the inequality is valid for the submodularity.
\end{IEEEproof}

The following result allows the manager to further restrict the class of mechanisms to take into consideration.

\begin{lemma}
There exists an optimal incentive compatible direct mechanisms such that, $\forall \, t \in T$, the randomized intervention rule sustains the suggested action profile without intervention in $\Gamma_t$.
\label{prop:kakutani}
\end{lemma}
\begin{IEEEproof}
See Appendix \ref{app:kakutani}
\end{IEEEproof}

%In this paper we consider randomized intervention rules because the restriction to pure intervention rules might lead to suboptimal results.
%However, Lemma \ref{prop:kakutani} states that this is not the case, we can restrict to pure intervention rules without loosing optimality. 
Lemma \ref{prop:kakutani} suggests the idea to decouple the original problem, \textbf{OICDM}, into two sub-problems.
First we can calculate the optimal suggested action profile $d^S(t)$ under the constraint that users adopting that action profile have incentives to report their real type.
Finally, it is sufficient to identify an intervention rule able to sustain $d^S(t)$ without intervention in $\Gamma_t$.
This is formalized in the following.

Consider the mechanism $\left(T, D, \overline{d}^S, \overline{\pi} \right) $, where
\begin{align}
& \overline{d}^S = \argmax_{d^S} \sum_{t \in T} P_t(t) U_0 \left( d_0^*, d^S(t), t \right) \nonumber\\
& \mbox{subject to:} \nonumber\\
&\sum_{t_{-i} \in T_{-i}} P_t(t \mid \tau_i) U_i \left( d_0^*, d^S(t_{-i}, \tau_i) , t \right) \geq \sum_{t_{-i} \in T_{-i}} P_t(t \mid \tau_i) U_i \left( d_0^*, d_{-i}^S(t_{-i}, \hat{\tau}_i), \hat{\delta}_i(d^S_i(t_{-i}, \hat{\tau}_i)), t \right) \nonumber\\
& \forall \, i \in \left\lbrace 1, ..., n \right\rbrace, \;\; \forall \, \tau_i \in T_i, \;\; \forall \, \hat{\tau}_i \in T_i, \;\; \forall \, \hat{\delta}_i : D_i \rightarrow D_i
\label{eq:subprob1}
\end{align}
and, $\forall \, t \in T$, %$\overline{\pi} \left(f \mid t \right)$ is equal to $1$ for a certain $f \in \mathcal{F}^{\overline{d}^S,t}$, $0$ otherwise.
\begin{equation}
\overline{\pi} \left( \cdot \mid t \right) \in \mathcal{F}^{\overline{d}^S,t}
%\overline{\pi} \left(f \mid t \right) =
%\left\lbrace
%\begin{array}{lll}
%1 & \mbox{for a certain} \; f \in \mathcal{F}^{\overline{d}^S,t} \\
%0 & \mbox{otherwise}
%\end{array}
%\right. %&\forall \, t \in T \; ,
\label{eq:subprob2}
\end{equation}

\begin{proposition}
The mechanism $\left(T, D, \overline{d}^S, \overline{\pi} \right) $ 
is an optimal incentive compatible direct mechanism.
\label{prop}
\end{proposition}
\begin{IEEEproof}
Eq. (\ref{eq:subprob2}) says that we are looking for a mechanism where, $\forall \, t \in T$, 
the randomized intervention rule sustains the suggested action profile without 
intervention in $\Gamma_t$. 
Moreover, the constraint of Eq. (\ref{eq:subprob1}) says that the users have the incentive
to reveal their true types if they adopt the suggested action profile.
Lemma \ref{prop:kakutani} states that such a class of mechanisms is optimal, hence, the solution of 
Eqs. (\ref{eq:subprob1})-(\ref{eq:subprob2}) gives an optimal incentive compatible direct mechanism.
\end{IEEEproof}

\begin{corollary}
The intervention capability $\left( D_0, \mathcal{F} \right)$ is optimal with respect to $\Gamma$.
\label{cor}
\end{corollary}

\section{Sub-optimal incentive compatible direct mechanisms}\label{sec:sdm}

In this Section we provide practical tools for the manager to design efficient coordination mechanisms.
Although we have characterized the optimal mechanism, other  schemes are also of interest, for several reasons.  
First of all, the optimal intervention scheme may be very difficult to compute, even in the decoupled version of 
Eqs. (\ref{eq:subprob1})-(\ref{eq:subprob2}).  
It is therefore of some interest to consider intervention schemes that are sub-optimal but easy to compute.
Moreover, in some situations, it may not be possible for the users to communicate with the manager, so it is natural to consider intervention schemes that do not require the users to make reports.
In the following, we address both issues.
In Subsection \ref{sec:sa} we describe an algorithm that converges to an incentive compatible direct mechanism where the recommended actions are as close as possible to the optimal ones.
In Subsection \ref{sec:oap} we consider a mechanism that is independent of users' reports.

\subsection{Algorithm that converges to an incentive compatible direct mechanism}\label{sec:sa}

In this Section we propose a general algorithm (see Algorithm \ref{algo1}) that converges to an incentive compatible direct mechanism.
Such algorithm is run by the intervention device at the beginning of the interaction with the users in order to obtain the mechanism to adopt.
After that, the interaction between the intervention device and the users is as usual: the intervention device communicates the mechanism, the users report their type, the intervention device suggests the actions to adopt, the users take actions, and finally the intervention device monitors users' actions and intervenes.
This algorithm can be applied when the suggested action profile, for every type profile $t$ and at each step of 
the algorithm, is sustainable  without intervention in $\Gamma_t$.
The suggested action profile will never be lower than the optimal action profile $d^*(t)$ and higher than the 
$NE$ action profile $d^{NE^0}(t)$ of $\Gamma^0_t$, so it is sufficient that $\mathcal{F}^{d,t}$ is non empty 
$\forall \, t \in T$ and $\forall \, d \in D$ so that $d^*(t) \leq d \leq d^{NE^0}(t)$. 

We denote by $W_i(t_i, \hat{t}_i)$ the expected utility that user $i$, with type $t_i$, obtains reporting type $\hat{t}_i$ and adopting the suggested action, assuming that the other users are honest and obedient, i.e.,
\begin{align}
W_i(t_i, \hat{t}_i) = \sum_{t_{-i} \in T_{-i}} \sum_{f \in \mathcal{F}} P_t(t \mid t_i) \pi \left( f \mid \hat{t} \right) U_i \left( f, d^S(\hat{t}), t \right)
\label{eq:1}
\end{align}
where we used the notation $\hat{t} = \left( t_1, \dots, t_{i-1}, \hat{t}_i, t_{i+1}, \dots, t_n \right) $.

\begin{figure}[p]%[htp]
     \centering
          \includegraphics[width=\figw]{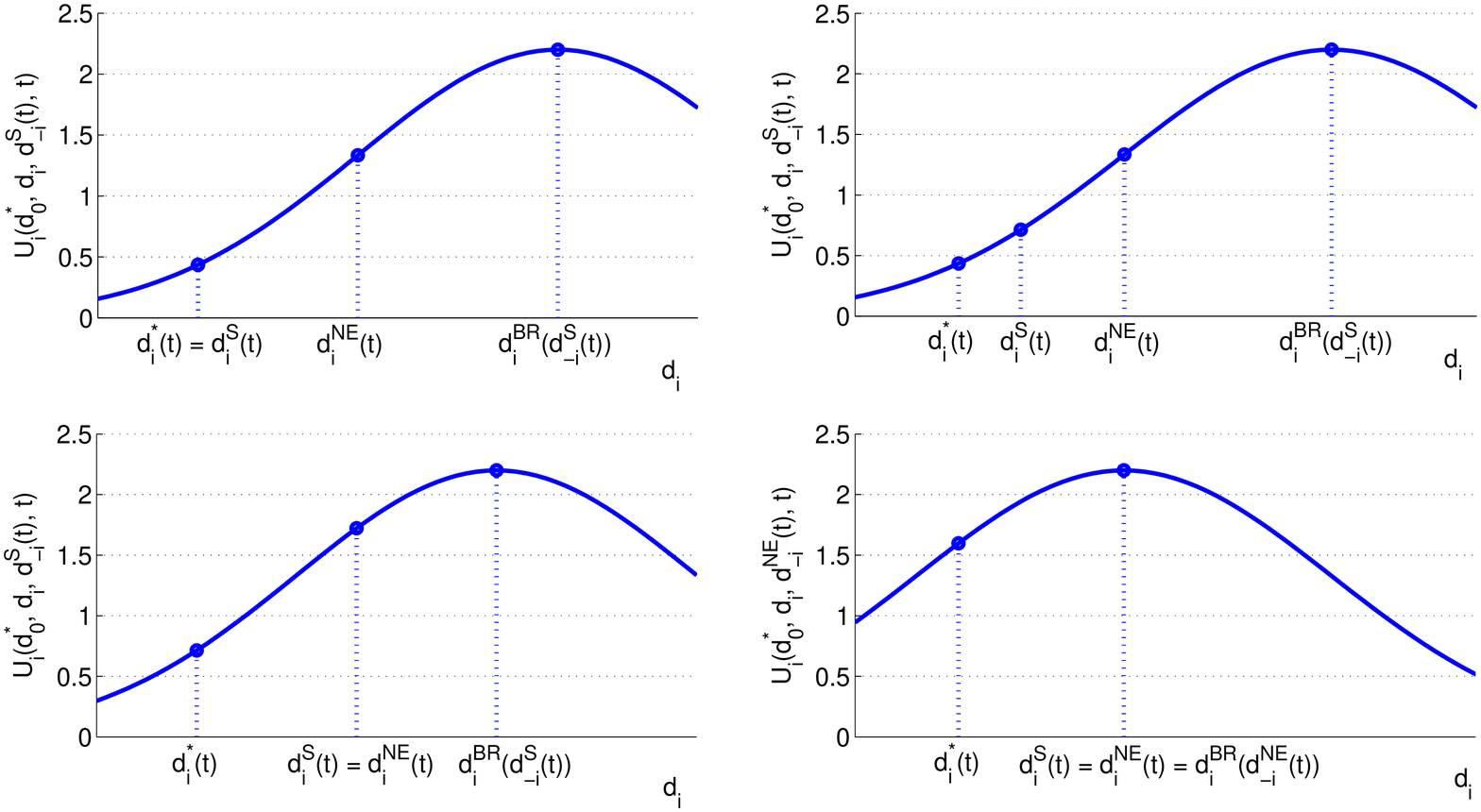}
\caption{User $i$'s utility vs. user $i$'s action, for different suggested actions}
\label{fig:algo_rep}
\end{figure}

The algorithm has been designed with the idea to minimize the distance between the optimal action profile $d^*(t)$ and the suggested action profile $d^S(t)$, for each possible type profile $t$.  
To explain the idea behind the algorithm we use Fig. \ref{fig:algo_rep}, where $i$'s utility is plotted with respect to $i$'s action, for a fixed type profile $t$ and assuming the other users adopt the suggested actions $d_{-i}^{S}(t)$.

The algorithm initializes the suggested action profile $d^S(t)$ equal to the optimal action profile $d^*(t)$ and selects a randomized intervention rule $\pi \left( \cdot \mid t \right)$ that sustains it without intervention, for every type profile $t \in T$. 
This situation is represented by the upper-left Fig. \ref{fig:algo_rep}.
Also the $NE$ and $i$'s best response action are represented, $d^{NE^0}(t)$ and $d^{BR}(d_{-i}^S(t))$.
By assumption \textbf{A6} $d^*(t) \leq d^{NE^0}(t)$ and by assumption \textbf{A5} $d^{NE^0}(t) \leq d^{BR}(d_{-i}^S(t)) $, because $d_{-i}^S(t) \leq d_{-i}^{NE^0}(t)$.
If $W_i(t_i, t_i) \geq W_i(t_i, \hat{t}_i)$, for every alternative $i$'s reported type $\hat{t}_i$, then user $i$ has an incentive to report its true type $t_i$. 
If, at a certain iteration of the algorithm, this is valid for all users and for all types they may have, then the algorithm stops and an incentive compatible direct mechanism is obtained.\footnote{Notice that, if a maximum efficiency incentive compatible direct mechanism exists, since it must satisfy the conditions of Proposition \ref{lemma1}, then the initialization of the algorithm corresponds to a maximum efficiency incentive compatible direct mechanism and the algorithm stops after the first iteration.} 

Conversely, suppose there exists a user $i$ and types $t_i$ and $\hat{t}_i$ such that 
$W_i(t_i, t_i) < W_i(t_i, \hat{t}_i)$, i.e., user $i$ has the incentive to report $\hat{t}_i$ when its type is $t_i$.
Then the suggested action $d_i^S(t)$ is increased by a quantity equal to $\epsilon_i$, 
moving it in the direction of the best response function $d_i^{BR}(d_{-i}^S(t))$, for every possible combination 
of types $t_{-i}$ of the other users, and updates the randomized intervention rule $\pi \left( \cdot \mid t \right)$ in order to sustain 
without intervention the new suggested action profile.
This has the effect, as represented by upper-right Fig. \ref{fig:algo_rep}, to increase $U_i \left( d_0^*, d^S(t), t \right)$, $\forall \, t_{-i} \in T_{-i}$, and therefore also the expected utility of $i$ when it has type $t_i$ and it is honest, $W(t_i, t_i)$.
This procedure is repeated as long as $W_i(t_i, t_i) < W_i(t_i, \hat{t}_i)$ and $d_i^S(t) \leq d_i^{NE^0}(t)$.
In case $i$'s suggested action $d_i^S(t)$ reaches $d_i^{NE^0}(t)$ and still $W_i(t_i, t_i) < W_i(t_i, \hat{t}_i)$, then the suggested action of user $k$, $d_k^S(t)$, is increased by a quantity equal to $\epsilon_k$, $\forall \, k \in \mathcal{N}$, $k \neq i$, $\forall \, t_{-i} \in T_{-i}$.
As we can see from lower-left Fig. \ref{fig:algo_rep}, this means to move the best response function $d_i^{BR}(d_{-i}^S(t))$ in the direction of the suggested action $d_i^S(t)$.
If $d_k^S(t)$ reaches $d_k^{NE^0}(t)$ as well, $\forall \, k \in \mathcal{N}$, then $d_{-i}^S(t)$ coincides with the best response function $d^{BR}(d_{-i}^S(t))$, as represented in the lower-right Fig. \ref{fig:algo_rep}.
In fact, by definition, the $NE$ is the action profile such that every user is playing its best response action against the actions of the other users.
Since $d_i^S(t)$ coincides with $d^{BR}(d_{-i}^S(t))$, $\forall \, t_{-i} \in T_{-i}$, user $i$ is told to play its best action for every possible combination of the types of the other users.
Hence, user $i$ cannot increase its utility reporting a different type $\hat{t}_i$, therefore the mechanism is incentive compatible. 

The algorithm stops the first time each user has the incentive to declare its real type.
Since at each iteration the suggested action profiles are increased by a fixed amount, the algorithm converges after a finite number of iterations.
The higher the steps $\epsilon_i$, $i \in \mathcal{N}$, the lower the convergence time of the algorithm.
On the other hand, the lower the steps, the closer the suggested action profile to the optimal one.\footnote{Notice that, 
since no assumption such as convexity is made for the manager's expected utility $V_0$, an action profile closer to the 
optimal one does not necessarily imply a better outcome for the manager.}

\begin{algorithm}
\caption{General algorithm.} \label{algorithm:gen_algo}
\begin{algorithmic}[1]
\STATE \textbf{Initialization}: $\forall \, t \in T$, $d^S(t) = d^*(t)$, $\pi \left( \cdot \mid t \right) \in \mathcal{F}^{d^S,t}$.
\STATE \textbf{For} each user $i \in \mathcal{N}$ and each couple of states $t_i, \hat{t}_i \in T_i$
\STATE ~~~\textbf{If} $W_i(t_i, t_i) < W_i(t_i, \hat{t}_i)$ 
\STATE ~~~~~~\textbf{If} $d_i^S(t_i, t_{-i}) < d_i^{NE^0}(t_i, t_{-i})$ for some $t_{-i} \in T_{-i}$
\STATE ~~~~~~~~~$d_i^S(t_i, t_{-i}) \leftarrow \min\left\lbrace d_i^S(t_i, t_{-i}) + \epsilon_i, \; d_i^{NE^0}(t_i, t_{-i})  \right\rbrace$, $\pi \left( \cdot \mid t \right) \in \mathcal{F}^{d^S,t}$,  $\forall \, t_{-i} \in T_{-i}$ 
\STATE ~~~~~~\textbf{Else}
\STATE ~~~~~~~~~$d_k^S(t_i, t_{-i}) \leftarrow \min\left\lbrace d_k^S(t_i, t_{-i}) + \epsilon_k, \; d_k^{NE^0}(t_i, t_{-i})  \right\rbrace$, $\pi \left( \cdot \mid t \right) \in \mathcal{F}^{d^S,t}$, $\forall k \in \mathcal{N}$, $k \neq i$, $\forall \, t_{-i} \in T_{-i}$
\STATE Repeat from $2$ until $3$ is unsatisfied $\forall \, i$, $t_i$, $t_{-i}$
\end{algorithmic}
\label{algo1}
\end{algorithm}

%%[p] mette le figure alla fine dell'articolo
%\begin{figure}[p]%[htp]
%     \centering
%          \includegraphics[width=\figw]{Algo_rep2.eps}
%\caption{User $i$'s utility vs. user $i$'s action, for different suggested actions}
%\label{fig:algo_rep}
%\end{figure}

\subsection{A priori direct mechanism}\label{sec:oap}

In this Subsection we consider a new type of mechanism, namely an \emph{a priori mechanism}, where users' reports do not play any role for the final outcome.
This is particularly useful in situations where it is not possible for the users to communicate with the manager. 
%In this case, since no reports are received, the manager is forced to consider only mechanisms where the suggested actions and the randomized intervention rule are independent of users' reports.
However, also for scenarios where users can send reports, an a priori mechanism might represent a good sub-optimal mechanism that is efficient and easy to compute.

\begin{defn}
$\left( T, D, d^S, \pi \right) $ is an a priori direct mechanism if it is a  direct mechanism and the suggested action profile $d^S$ and the selected randomized intervention rule $\pi$ do not depend on users' reports.
$\left( T, D, d^S, \pi \right) $ is an a priori incentive compatible direct mechanism if it is an a priori direct mechanism and it is incentive compatible.
\end{defn}

In an a priori direct mechanism stages \textbf{1}-\textbf{3} described in Subsection \ref{sec:cm} can be compressed in only one stage in which the intervention device communicates to the users the suggested action profile $d^S$ and the randomized intervention rule $\pi$. 
In an a priori incentive compatible direct mechanism the incentive compatibility condition must be checked only for users' actions and \textbf{OICDM} simplifies in\footnote{Notice that the optimal a priori incentive compatible direct mechanism attainable solving (\ref{eq:in_comp3}) is in general suboptimal compared to the optimal a priori direct mechanism.
In fact, the revelation principle does not hold for a priori mechanisms since we are adding an additional constraint, forcing the mechanism to be independent of users' reports.}
\begin{align}
&\argmax_{d^S, \pi} \sum_{t \in T} \sum_{f \in \mathcal{F}} P_t(t) \pi \left( f \right) U_0 \left( f, d^S, t \right) \nonumber\\
& \mbox{subject to:} \nonumber\\
&\pi \left( f \right) \geq 0 \;\; , \;\; \sum_{x \in \mathcal{F}} \pi \left( x \mid t \right) = 1 \;\; , \;\; \forall f \in \mathcal{F} \nonumber\\
&\sum_{t_{-i} \in T_{-i}} \sum_{f \in \mathcal{F}} P_t(t \mid \tau_i) \pi \left( f \right) U_i \left( f, d^S, t \right) \geq \sum_{t_{-i} \in T_{-i}} \sum_{f \in \mathcal{F}} P_t(t \mid \tau_i) \pi \left( f \right) U_i \left( f, d_{-i}^S, \hat{\delta}_i(d^S_i), t \right) \nonumber\\
& \forall \, i \in \left\lbrace 1, ..., n \right\rbrace, \;\; \forall \, \tau_i \in T_i, \;\; \forall \, \hat{\delta}_i : D_i \rightarrow D_i
\label{eq:in_comp3}
\end{align}

\begin{defn}
A randomized intervention rule $\pi$ sustains an action profile $d \in D$ in $\Gamma$ if $d$ is a $BNE$ of the game $\Gamma$, i.e., if, $\forall \, i \in \mathcal{N} \; , \; \forall \, \tau_i \in T_i \; , \; \forall \, \tilde{d}_i \in D_i$,
\begin{align}
\sum_{t_{-i} \in T_{-i}} \sum_{f \in \mathcal{F}} P_t(t \mid \tau_i) \pi \left( f \mid t \right) U_i \left( f, d, t \right) \geq \sum_{t_{-i} \in T_{-i}} \sum_{f \in \mathcal{F}} P_t(t \mid \tau_i) \pi \left( f \mid t \right) U_i \left( f, \tilde{d}_i, d_{-i}, t \right) % \;\; , \;\; \forall \, i \in \mathcal{N} \; , \; \forall \, \tau_i \in T_i \; , \; \forall \, \tilde{d}_i \in D_i 
\label{eq:sus_gamma}
\end{align}

A randomized intervention rule $\pi$ sustains an action profile $d \in D$ in $\Gamma_t$ without intervention if $\pi$ sustains $d$ and $f(d) = d_0^*$ for every intervention rule $f$ such that $\pi \left( f \mid t \right) > 0$. 
If such $\pi$ exists, we say that $d$ is sustainable without intervention.

%A \emph{pure} intervention rule $\hat{f} \in \mathcal{F}$ sustains an action profile $d \in D$ in $\Gamma_t$ (without intervention) if $\hat{\pi}$ sustains $d$ in $\Gamma_t$ (without intervention), where $\hat{\pi} \left( f \mid t \right) = 1$ if $f = \hat{f}$, $0$ otherwise.

\end{defn}

If any action profile is sustainable without intervention in $\Gamma$, then (\ref{eq:in_comp3}) can be decoupled and an optimal a priori incentive compatible direct mechanism can be computed as a solution to the following unconstrained optimization problem: 
\begin{equation}
\overline{d}^S = \argmax_{d^S} \sum_{t \in T} P_t(t) U_0 \left( d_0^*, d^S, t \right) 
\label{eq:micky}
\end{equation}
and $\overline{\pi} \left(f \right)$ sustains $\overline{d}^S$ in $\Gamma$ without intervention.
%\begin{align}
%&\overline{\pi} \left(f \right) =
%\left\lbrace
%\begin{array}{lll}
%1 & \mbox{for a certain} \; f \in \mathcal{F}^{\overline{d}^S,t} \\
%0 & \mbox{otherwise}
%\end{array}
%\right. 
%\label{eq:mouse}
%\end{align}

%[p] mette le figure alla fine dell'articolo

%[p] mette le figure alla fine dell'articolo

% LUCA: HO SISTEMATO TUTTA LA PRIMA PARTE E GLI APPENDICI IN LOGICA. ORA DEVO SISTEMARE LA SECONDA PARTE CHE RICHIEDE UN PO' PIU' DI LAVORO

\section{Application to flow control}\label{sec:fcg}

In this Section we apply the results derived in Sections \ref{sec:dm} and \ref{sec:sdm} for the 
abstract framework to a concrete scenario: the design of a flow control management system.

\subsection{Formulation of the flow control problem}\label{sec:mdl}

We consider $n$ Poisson streams of packets with arrival rates $d_1$, $d_2$, ..., $d_n$ that are serviced by a single server with exponentially distributed service times with mean $\frac{1}{\mu}$.
Since we assume that all packets have the same length, we will talk interchangeably of arrival rate ($\frac{pkt}{s}$) and transmission rate ($Mbps$), and $\mu$ can be seen as the channel capacity, in $\frac{pkt}{s}$, after the server.\footnote{We consider packets of the same length to keep a simple notation and because the qualitative results are not affected by this hypothesis. However, the model and the analysis can be easily extended to take into account packets of different lengths.} 
We refer to each stream of packets as a user. 
We assume that each user $i$ can control its own traffic (e.g., by adjusting the coding quality of its communication), i.e., it can select its transmission rate $d_i \in D_i = \left[0, \; \mu \right]$. 
As represented by Fig. \ref{fig:flow}, the system is an M/M/1 queue with an input arrival rate $\lambda = \sum_{i=1}^n d_i$.

\begin{figure}%[htp]
     \centering
          \includegraphics[width=\figw]{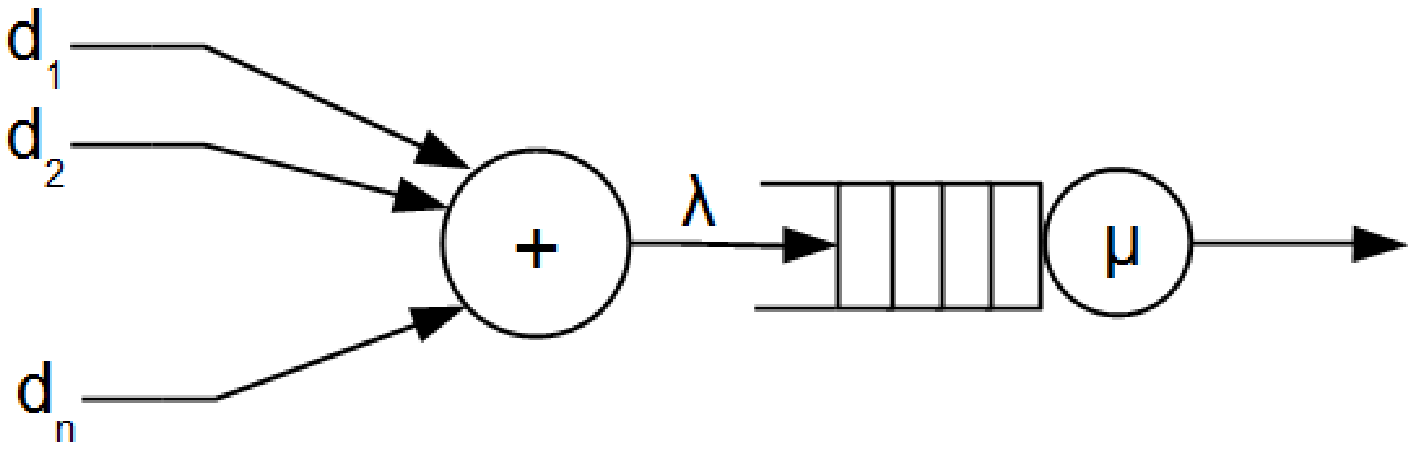}
\caption{Representation of a flow control application as a M/M/1 queue}
\label{fig:flow}
\end{figure}

In most cases a user is faced with two conflicting objectives, i.e., to maximize its throughput\footnote{Here the throughput refers to the traffic the server is able to service, i.e., the transmission rate available to the user, and does not take into account the packets lost due to physical layer transmission errors.} and to minimize its average delay.
The conflict between throughput and delay is obvious since as more traffic enters the server queue the delays become larger.
In order to incorporate these two measures in a single performance metric, the concept of power has been proposed in \cite{Giessler} and later extended in \cite{Kleinrock}.
It is defined as the ratio between the throughput and the average delay, where the exponent of the throughput is a positive constant.
We can therefore write $i$'s utility as
\begin{align}
U_i(d, t_i) = d_i^{t_i} \left( \mu - \lambda \right) = d_i^{t_i} \left( \mu - \sum_{i=1}^n d_i \right)
\label{eq:ui}
\end{align}
where $d = \left( d_1, \dots, d_n \right) $ denotes the transmission rate profile and the parameter $t_i > 0$ represents user $i$'s type.

The value of $t_i$ may depend, for example, on the quality of service of the application corresponding 
to the $i$-th stream of packets. 
As we will see in Eqs. (\ref{eq:opt}) and (\ref{eq:NE}),
both considering compliant users and strategic users, 
the rate adopted by a user is increasing in its type.
This consideration suggests the idea that the higher the type of a user, the higher the importance of the rate, 
with respect to the delay, for that user.
As an example, streams of packets associated to delay dependent applications should have a low type 
while streams of packets associated to delay tolerant applications should have a high type.

In general, the applications a server has to deal with may change over time. 
For this reason it is useful to define a common type set $T_i = \left\lbrace \tau_1, \tau_2, ...,\tau_m \right\rbrace$, 
$m \in \mathbb{N}$, $\tau_k \in \mathbb{R}$, $\tau_1 < \tau_2 < ... < \tau_m$, 
$\forall i \in \mathcal{N} = \left\lbrace 1, 2, ..., n \right\rbrace $, whose elements represent all the possible 
types of users a server has to deal with.
Suppose that at the beginning of the communication a user does not know the types of the other users and the 
intervention device itself does not know the types of the users.
We assume that a common probability distribution exists and that user types are independent and identically 
distributed (i.i.d.) with $P(t_i)$ denoting the probability that a user has type 
$t_i$, $t_i \in T_i$, and $P_t(t) = \prod_{i=1}^n P(t_i)$ the probability that the type profile is $t$, $t \in T = T_i^n$.
$P(t_i)$ can be thought as the average fraction of applications having type $t_i$ that require services to the server.
%We use the notation $d_{-i}$ and $t_{-i}$ to denote the users' actions and the users' types except for user $i$, while $D_{-i}$ and $T_{-i}$ denote the sets of all possible combinations of users' actions and users' types except for user $i$.

The network must be designed to operate efficiently following the manager's objective, which can be quantified by a utility function. 
We assume that the manager's utility is the geometric mean of the users' utilities:
\begin{align}
U_0(d, t) = \sqrt[n]{\prod_{i=1}^n U_i^+(d, t_i)} = \left( \mu - \lambda \right)^+ \prod_{i=1}^n d_i^{\frac{t_i}{n}} 
\label{eq:u0}
\end{align}
where $\left( x \right)^+ = \max \left\lbrace x \; , \; 0 \right\rbrace $.\footnote{
%In the definition of intervention device utility 
We consider $U_i^+$ instead of $U_i$ for mathematical reasons, because utilities as defined in Eq. (\ref{eq:ui}) 
may also be negative, and the geometric mean would lose meaning with negative quantities. 
Anyway, notice that it is in the self interest of both the users and the manager to have $\lambda \leq \mu$, i.e., working 
in the sub-space of the original domain such that $U_i^+=U_i$.}
This choice allows to maintain a balance between two competing interests a benevolent manager might have: to 
maximize the social welfare of the network (defined as the sum utility) and to allocate resources fairly, 
giving to users similar utilities. 
Notice that maximizing $U_0(d, t)$ with respect to users' actions is equivalent to maximizing a proportional 
fairness of users' utilities, i.e., $\sum_{i=1}^n \ln U_i^+(d, t_i)$, and the optimal solution 
$d^* = \left\lbrace d_i^* \right\rbrace_{i=1}^n$ is given by (see \cite{YiMihaela_TCOM})
\begin{align}
d_i^*(t) = \dfrac{t_i \mu}{n + \sum_{k=1}^n t_k} %\;\; , \;\; i \in \mathcal{N}
\label{eq:opt}
\end{align} 
%which is increasing in $i$'s type.

We denote by $V_i(d(t), t_i)$ and $V_0(d(t))$ the expected (with respect to the types) utilities of user $i$ having type $t_i$ and of the manager, where $d(t)$ represents the action adopted by the users when the type profile is $t$, i.e., 
\begin{align}
V_i(d(t), t_i) = \sum_{t_{-i} \in T_{-i}} P_t(t \mid t_i) U_i(d(t), t)  
\label{eq:vi_23}
\end{align}
\begin{align}
V_0(d(t)) = \sum_{t \in T} P_t(t) U_0(d(t), t)  
\label{eq:v0_23}
\end{align}

Hence, the maximum expected utility that the manager can obtain when users are compliant is $V_0^{ME} = V_0(d^*(t))$.

\subsection{The flow control games}

In this Subsection we compute the outcome of a flow control problem considering self-interested and strategic users, for both the complete and the incomplete information scenarios.
Moreover, we quantify the loss of efficiency of the manager's utility with respect to the maximum efficiency utility.

\subsubsection{The complete information game $\Gamma^0_t$}

We define the complete information game 
\begin{equation}
\Gamma^0_t = \left( \mathcal{N}, D, \left\lbrace U_i(\cdot, t) \right\rbrace_{i=1}^N  \right)
\label{eq:gamma_0_1}
\end{equation}
where each user $i$ selects its action $d_i(t)$ strategically, knowing the types $t$ of all the users.

The unique $NE$ $d_i^{NE^0}(t)$ of $\Gamma^0_t$ is, $\forall \, i \in \mathcal{N}$, (see \cite{YiMihaela_TCOM})
\begin{align}
d_i^{NE^0}(t) = \dfrac{t_i \mu}{1 + \sum_{k=1}^n t_k} %\;\; , \;\; i \in \mathcal{N}
\label{eq:NE}
\end{align}
Notice that strategic users use the resources more heavily with respect to compliant users, i.e., $d_i^{NE^0}(t) > d_i^*(t)$, $\forall \, i \in \mathcal{N}$ and $\forall \, t \in T$ (excluding the trivial case $n=1$).

The manager's expected utility in the complete information scenario is equal to $V_0(d^{NE^0}(t))$.
%\begin{align}
%V_0^{CI} = \sum_{t \in T} P_t(t) U_0(d_0^*, d^{NE^0}(t), t)  
%\label{eq:v0_1}
%\end{align}

\subsubsection{The Bayesian game $\Gamma^0$}\label{sec:pippo}

We define the incomplete information game  
\begin{equation}
\Gamma^0 = \left( \mathcal{N}, D, T, P_t, \left\lbrace U_i(\cdot, t) \right\rbrace_{i=1}^n  \right)
\label{eq:gamma_0}
\end{equation}
where each user $i$ selects its action $d_i(t_i)$ strategically, knowing its own type $t_i$ and the probability distribution over the types of the other users, $P_t$.

\begin{proposition}
There exists a unique Bayesian Nash Equilibrium $d^{BNE}(t)$ of $\Gamma^0$ which can be obtained by solving a linear system $\mathbf{A} d^{BNE} = b$. In addition, the inverse of $\mathbf{A}$, $\mathbf{A}^{-1}$, can be computed analytically.\footnote{The expressions of $b$, $\mathbf{A}$ and $\mathbf{A}^{-1}$ can be found in Appendix \ref{app:matrix_inv}.}
\label{prop:BNE}
\end{proposition}

\begin{IEEEproof}
See Appendix \ref{app:matrix_inv}.
\end{IEEEproof}

The manager's expected utility in the incomplete information scenario is equal to $V_0(d^{BNE}(t))$.
%\begin{align}
%V_0 = \sum_{t \in T} P_t(t) U_0(d_0^*, d^{BNE}(t), t)  
%\label{eq:v0_2}
%\end{align}

\subsubsection{Results}

Fig. \ref{fig:flow} shows the manager's expected utility with respect to the number of users, considering $\mu = 5 \, Mbps$ and a type set $T_i = \left\lbrace 0.1, \, 1 \right\rbrace $ with uniformly distributed types.
The upper curve represents the maximum efficiency utility, attainable when users are compliant to the manager, 
while the dashed and the dotted lines represent the manager's utility when users are strategic in the complete and 
incomplete information cases respectively.
%This can be considered an upper bound of the performances achievable. 
The manager's utility when users act strategically, both for the complete and incomplete information scenarios, is far below compared to the maximum efficiency utility.
Notice that the manager can obtain a higher utility in the incomplete information scenario with respect to 
the complete information scenario, at least when there are more than three users in the system.
This agrees with the results of \cite{Holmstrom1980, CrawSobel} where, in a strategic setting, 
the less closely related the agents' goals the lower the quantity of information they prefer to exchange.
In our case, the objective of the manager becomes less closely related to the objective of a single user as 
the number of total users increases. 
In fact, the manager's objective is to increase the utility of all users in a fair way, while the goal of a user 
is to improve only its own utility, at the cost of the utility of all the other users.
Hence, as the number of users increases, the selfishness of a single user has a higher negative impact on the 
manager's objective.

\begin{figure}%[htp]
     \centering
          \includegraphics[width=\figw]{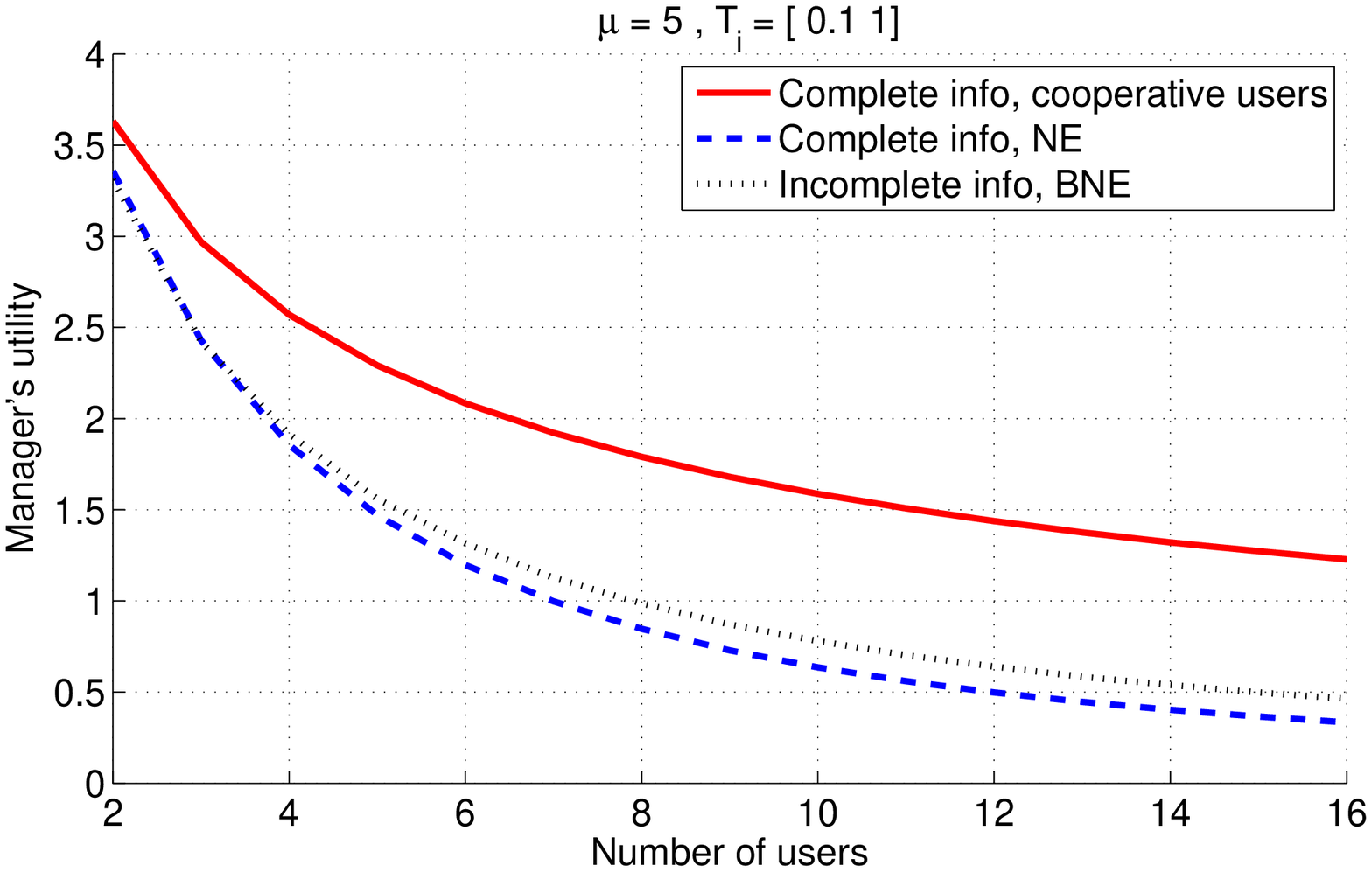}
\caption{Manager's utility as a function of the number of users}
\label{fig:flow}
\end{figure}

\subsection{The flow control games with intervention}

Fig. \ref{fig:flow} shows that the manager's expected utility in strategic settings is much lower that the manager's expected utility in cooperative settings, when users are compliant.
Here we ask whether the manager can design the system in order to make it robust against self-interest strategic users, filling, at least partially, the gap between the maximum efficiency utility and the manager's expected utility in strategic settings.

Using the same notations as in Section \ref{sec:frm}, we consider an intervention device that transmits a stream of packets to the server with a rate $d_0 \in D_0 = \left[0, \; d_0^M \right]$, following an intervention rule $f: D \rightarrow D_0$.
$d_0^M$ represents the maximum rate the intervention device is able to transmit. 
The framework introduced is Subsection \ref{sec:mdl} can be easily modified to take into account the effect of the intervention device, that increases the incoming traffic of the server $\lambda = \sum_{i=1}^n d_i + f(d)$.
The users' and the manager's utilities change accordingly:
\begin{align}
U_i(f, d, t_i) &= d_i^{t_i} \left( \mu - \lambda \right) = d_i^{t_i} \left( \mu - \sum_{i=1}^n d_i - f(d) \right) \nonumber\\
U_0(f, d, t) &= \sqrt[n]{\prod_{i=1}^n U_i^+(f, d, t_i)} = \left( \mu - \lambda \right)^+ \prod_{i=1}^n d_i^{\frac{t_i}{n}} 
\label{eq:ui}
\end{align}

It is straightforward to check that the users' and the manager's utilities satisfy assumptions \textbf{A1}-\textbf{A6}.
In particular, the manager's preferred action is $d_0 = 0$ (i.e., no intervention), and the game $\Gamma^0_t$ 
defined in Subsection \ref{sec:ut} coincides with the game $\Gamma^0_t$ defined in Subsection \ref{sec:pippo}.

%$U_i(d_0^*, d, t)$ is log-concave in $d_i$: 
%\begin{align}
%\dfrac{\partial \ln U_i(d_0^*, d, t)}{\partial d_i} &= \dfrac{t_i}{d_i} - \dfrac{1}{\mu - \sum_{k=1}^n d_k} \nonumber\\
%\dfrac{\partial^2 \ln U_i(d_0^*, d, t)}{\partial d_i^2} &= \dfrac{t_i}{d_i^2} - \dfrac{1}{\left( \mu - \sum_{k=1}^n d_k \right)^2 } < 0
%\label{eq:A4}
%\end{align}
%The best response function $d_i^{BR}(d_{-i}, t)$ is unique:
%\begin{align}
%d_i^{BR}(t) = \dfrac{t_i \left( \mu - \sum_{k=1, k\neq i}^n d_k\right) }{1 + t_i} \;\;\; i \in \mathcal{N}
%\label{eq:BR}
%\end{align}
%The game is submodular:
%\begin{align}
%\dfrac{\partial^2 \ln U_i(d_0^*, d, t)}{\partial d_i \partial d_j} =  - \dfrac{1}{\left( \mu - \sum_{k=1}^n d_k \right)^2 } < 0
%\label{eq:sub}
%\end{align}
%And the unique NE of $\Gamma^0_t$ is \cite{YiMihaela_TCOM}
%\begin{align}
%d_i^{NE^0}(t) = \dfrac{t_i \mu}{1 + \sum_{k=1}^n t_k} \;\;\; i \in \mathcal{N}
%\label{eq:NE}
%\end{align}
%is such that $d_i^{NE^0}(t) > d_i^*(t)$, $\forall i \in \mathcal{N}$ (excluding the trivial case $n=1$).

In the following we define a simple class of intervention rules, the class of affine intervention rules, where the intervention level increases linearly with the users' actions. 
We limit the design of intervention rules to affine intervention rules, i.e., $\mathcal{F}$ coincides with the class of affine intervention rules.
It may seem restrictive to constrain the intervention device to such a simple class of intervention rules. 
However, under certain conditions, the class of affine intervention rules will turn out to be optimal, i.e., it is not possible to increase the manager's utility by expanding the intervention rule set $\mathcal{F}$.

\begin{defn}
$f: D \rightarrow D_0$ is an affine intervention rule if 
\begin{align}
f(d) = \left[ \sum_{i=1}^n c_i (d_i - \tilde{d}_i) \right]_0^{d_0^M}
\label{eq:con}
\end{align}
for certain parameters $\tilde{d}_i \geq 0$ and $c_i \geq 0$, where $\left[ \cdot \right]_a^b = \min \left\lbrace \max \left\lbrace a, \cdot \right\rbrace, b \right\rbrace $.
\end{defn}

In an affine intervention rule, $\tilde{d}_i$ represents a target action for user $i$ while $c_i$ represents the 
rate of increase of the intervention level due to an increase of $i$'s action.
If the action profile $d$ is lower than or equal to the target action profile 
$\tilde{d} = \left( \tilde{d}_1, \cdots, \tilde{d}_n \right) $, then the intervention level is equal to $0$.
If the intervention level is higher than $0$, then some user is adopting an action higher then the target one. 
In this case, an increase by an amount $\epsilon$ of $i$'s action causes an increase in the intervention 
level by an amount $c_i \epsilon$. 

\begin{figure}%[htp]
     \centering
          \includegraphics[width=\figw]{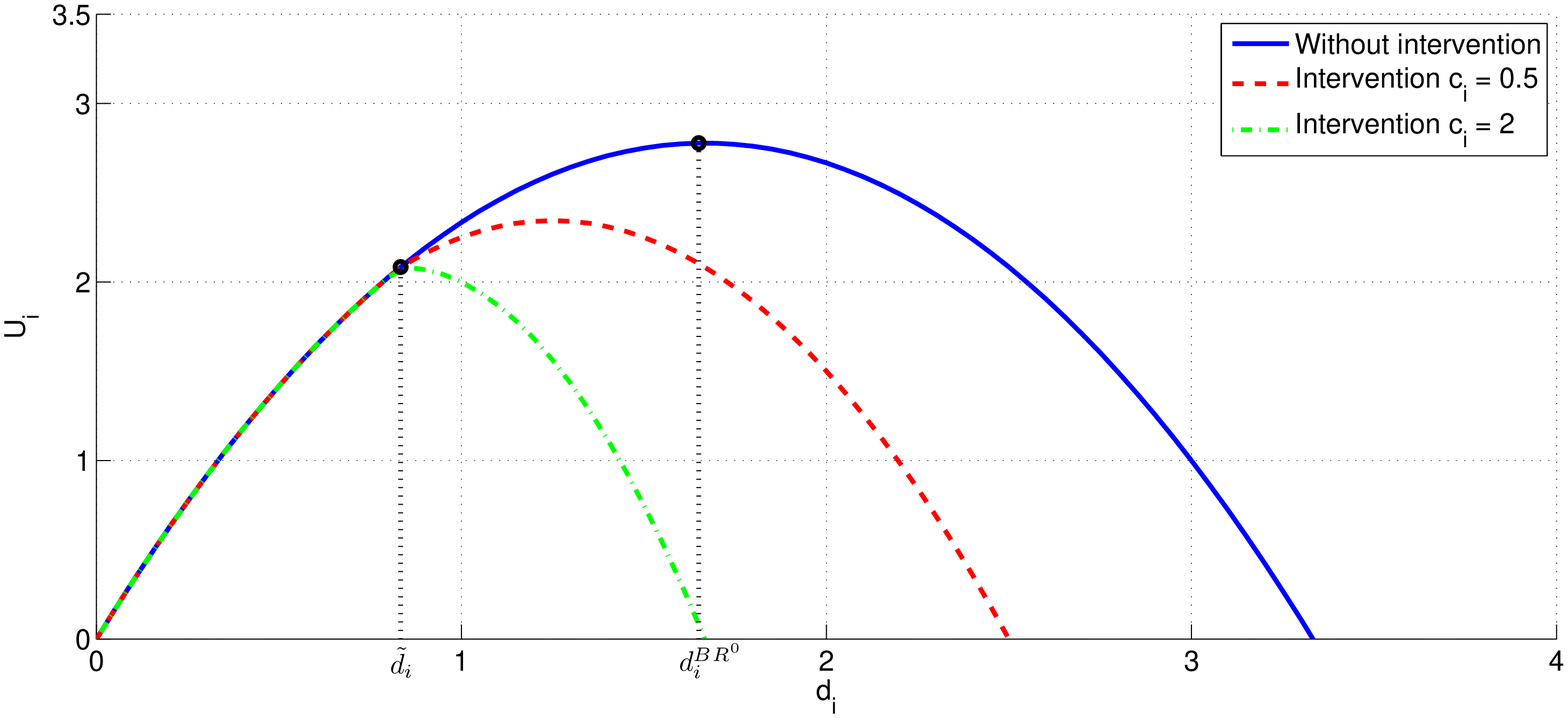}
\caption{User $i$'s utility as a function of user $i$'s action for different intervention rules}
\label{fig:aff_rule}
\end{figure}

Fig. \ref{fig:aff_rule} shows how an affine intervention rule changes the relation between $i$'s utility and $i$'s action.
The utility of user $i$ is plotted for tree cases: assuming that the intervention device never intervenes and 
assuming that the intervention device adopts a linear intervention rule, for two different values of the parameter $c_i$. 
We consider that the other users adopt the target action profile.
For an action $d_i$ lower than the target action $\tilde{d}_i$, $i$'s utility is as if the intervention device 
did not exist. 
However, for an action $d_i$ higher than the target action $\tilde{d}_i$, $i$'s utility is lower compared to the 
utility it would have obtained without intervention device, and the gap increases as $c_i$ increases.

In the following, we provide the tools for the manager to design the intervention rule, for both the complete and the incomplete information scenarios.

\subsubsection{The complete information game $\Gamma_t$}\label{sec:comp_info}

This is the scenario considered in Subsection \ref{sec:ci}, where the users adopt actions strategically, knowing the type of the other users and the intervention rule.
The interaction between users is modeled with the game $\Gamma_t$,
\begin{equation}
\Gamma_t = \left( \mathcal{N}, D, \left\lbrace U_i(f, \cdot, t_i) \right\rbrace_{i=1}^n \right),
\label{eq:gamma_t}
\end{equation}
where the utilities $U_i$, $i \in \mathcal{N}$, are calculated as in Eq. (\ref{eq:ui}).
The outcome of such interaction is represented by the $NE$. 

The manager faces the problem of designing a randomized intervention rule so that there exists a $NE$ of the game $\Gamma_t$ that gives it the highest utility among what is achievable with all possible $NEs$.
We will see that it can reach this objective considering only \emph{pure} intervention rules, which are simpler to implement than randomized intervention rules.

%The solution concept to adopt in this situation is the $NE$ and the intervention device has to select a randomize intervention rule in order to obtain the $NE$ that maximizes its utility
%\begin{equation}
%\overline{U}_0 \left( d, t \right) = \mathbb{E}_f \left[ U_0 \left( f, d, t), t \right) \right] =
%\sum_{f \in \mathcal{F}} \pi \left( f \mid t \right) U_0 \left( f, d, t), t \right). 
%\label{eq:U_cs_0}
%\end{equation}
%
%
%Assuming intervention device will select a certain affine intervention rule with probability $1$, we can rewrite utility of $\Gamma_t$ as
%\begin{equation}
%\overline{U}_i \left( d, t \right) =  U_i \left( f, d, t \right) =  d_i^{t_i} \left( \mu - \sum_{k=1}^n d_k - \left[ \sum_{i=1}^n c_i (d_i - \tilde{d}_i) \right]_0^{d_0^M} \right)
%\label{eq:U_i_over_1}
%\end{equation}

\begin{lemma}
Consider the affine intervention rule $f$ such that, for every user $i \in \mathcal{N}$,
\begin{align}
c_i \geq \dfrac{t_i \left( \mu - \sum_{k=1}^n \tilde{d}_k \right) - \tilde{d}_i}{\tilde{d}_i} \;\;\; , \;\;\; d_0^M \geq \dfrac{c_i \left[ t_i \left( \mu - \sum_{k=1}^n \tilde{d}_k \right) - \tilde{d}_i \right] }{1 + t_i (1 + c_i)} 
\label{eq:conditions}
\end{align}

If $\tilde{d} \leq d^{NE^0}$, then $f$ sustains $\tilde{d}$ in $\Gamma_t$ without intervention.
\label{lemma_interv}
\end{lemma}

\begin{IEEEproof}
See Appendix \ref{app:interv_NE}
\end{IEEEproof}

\textit{Interpretation:} Selecting a $c_i$ high enough and if the intervention device is able to transmit with a large enough transmission rate, the threat of punishment discourages the users from adopting actions higher than the target. 
This situation is shown in Fig. \ref{fig:aff_rule} for $c_i = 2$. 
Hence, if the utility of user $i$ is increasing before the target action $\tilde{d}_i$ (in particular, this is valid if $\tilde{d}_i \leq d^{NE^0}_i$), as in Fig. \ref{fig:aff_rule}, the target action $\tilde{d}_i$ becomes the best response action 
for user $i$.

\begin{proposition} 

If $d_0^M \geq \dfrac{\mu}{1+\tau_1}$, then the optimal strategy profile $d^*(t)$, $\forall t \in T$, 
is sustainable without intervention using an affine intervention rule $f$ with $\tilde{d} = d^*(t)$ and $c_i \geq n-1$.

If $d_0^M \geq \mu$, then every strategy profile $d \leq d^{NE^0}(t)$, $\forall t \in T$, is sustainable without 
intervention using an affine intervention rule $f$ with a high enough $c_i$, $i \in \mathcal{N}$.

% \begin{itemize}
% \item if $d_0^M \geq \mu \left( 1 - \dfrac{\tau_1}{1+\tau_1}\right) $, then the optimal strategy profile $d^*(t)$, $\forall t \in T$, is sustainable without intervention using an affine intervention rule with $\tilde{d} = d^*(t)$ and $c_i \geq n-1$.
% \item if $d_0^M \geq \mu$, then every strategy profile $d \leq d^{NE^0}(t)$, $\forall t \in T$, is sustainable without intervention using an affine intervention rule with a high enough $c_i$, $i \in \mathcal{N}$.
% \end{itemize}
\label{sustain}
\end{proposition}

\begin{IEEEproof}
First, consider the second affirmation.
The condition of Eq. (\ref{eq:conditions}) on $d_0^M$ is automatically satisfied if the right hand side is lower than $0$.
Moreover, if it is higher than $0$, the right hand side is increasing in $c_i$. 
In fact, the function $g(c_i) = \frac{a c_i}{b + d c_i}$, with $a,b \geq 0$, is increasing in $c_i$, because $g'(c_i) = \frac{a b}{(b + d c_i)^2} > 0$.
Thus, the condition of Eq. (\ref{eq:conditions}) on $d_0^M$ becomes stricter as $c_i$ increases.
Taking the limit for $c_i \rightarrow + \infty$ we can find the following stricter condition on $d_0^M$ that does not depend on $c_i$:
\begin{align}
d_0^M \geq \dfrac{ t_i \left( \mu - \sum_{k=1}^n \tilde{d}_k \right) - \tilde{d}_i }{t_i} = \mu - \sum_{k=1}^n \tilde{d}_k - \dfrac{\tilde{d}_i \left( t_i + 1 \right) }{t_i}
\end{align}
In order to obtain conditions that are independent of users' types and action profiles to sustain, we can consider the following stricter conditions:
\begin{align}
d_0^M \geq \mu - \sum_{k=1}^n \tilde{d}_k - \dfrac{\tilde{d}_i \left( \tau_m + 1 \right) }{\tau_m} \;\;\; , \;\;\; d_0^M \geq \mu 
\label{eq:d_0_cond}
\end{align}
As for $c_i$, we can find a stricter condition independent of users' types substituting  $t_i$ with $\tau_m$.
Thus, once the action profile to sustain is fixed, it is sufficient to select a $c_i$ satisfying
\begin{equation}
c_i \geq \dfrac{\tau_m \left( \mu - \sum_{k=1}^n \tilde{d}_k \right) - \tilde{d}_i}{\tilde{d}_i}
\label{eq:ultimo}
\end{equation}

Now consider the first affirmation.
Substituting $d^*(t)$ into Eq. (\ref{eq:conditions}) we obtain
\begin{align}
c_i \geq \dfrac{t_i \left( \mu - \sum_{k=1}^n \tilde{d}_k \right)}{\tilde{d}_i} - 1 = n + \sum_{k=1}^n t_k - \dfrac{t_i \mu \sum_{k=1}^n t_k}{t_i \mu} - 1 = n-1
\end{align}
As to $d_0^M$, substituting $d^*(t)$ into the first condition of Eq. (\ref{eq:d_0_cond}) we obtain
\begin{align}
d_0^M \geq \mu - \mu \dfrac{\sum_{k=1}^n t_k}{n + \sum_{k=1}^n t_k} = \dfrac{n \mu}{n + \sum_{k=1}^n t_k}
\end{align}
Finally, since the right hand side is decreasing in $\sum_{k=1}^n t_k$, a stricter condition can be obtained substituting $t_k = \tau_1$, $\forall k \in \mathcal{N}$, obtaining
\begin{align}
d_0^M \geq \dfrac{\mu}{1+\tau_1} 
\end{align}
\end{IEEEproof}

If the intervention device is able to transmit a stream of packets with a rate higher than a certain threshold (that is upper-bounded by $\mu$), 
the manager can extract the maximum utility from the game $\Gamma_t$ adopting \emph{pure} affine intervention rules.
The following corollary is an implication of this consideration.

\begin{corollary}
If $d_0^M \geq \dfrac{\mu}{1+\tau_1}$, then the class of affine intervention rules is optimal with respect to $\Gamma_t$.
\end{corollary}

Finally, the manager's expected utility for the complete information scenario with intervention device, 
considering affine intervention rules and assuming the condition on $d_0^M$ is satisfied, is equal to the 
maximum efficiency utility $V_0(d^*(t))$.

\subsubsection{The Bayesian game $\Gamma$}\label{sec:bg}

This is the scenario considered in Subsection \ref{sec:cm}, where the users adopt actions strategically, knowing their own type and the probability distribution over the types of the other users.
The manager faces the problem of designing a direct mechanism $\left( T, D, d^S, \pi \right)$ so that the outcome of the system maximizes its own utility.
The interaction between users is modeled with the game $\Gamma$ defined by Eq. (\ref{eq:gamma}).
Since assumptions \textbf{A1}-\textbf{A6} are satisfied, we exploit the results obtained in Section \ref{sec:dm} 
%and \ref{sec:comp_info} 
for the abstract framework.

%First, we derive sufficient conditions on the type set $T_i$ that guarantees the existence of a maximum efficiency incentive compatible direct mechanism and we compute it. 
%
%Then we characterize an optimal incentive compatible direct mechanism in general.
%
%Finally, we propose two suboptimal solutions, in case the sufficient conditions are not satisfied.
%The first is the based on a priori direct mechanism, defined in subsection \ref{sec:oap}.
%The second is obtained applying the algorithm that converges to a suboptimal incentive compatible direct mechanism, as described in subsection \ref{sec:sa}.
%In general, the intervention device can compute both suboptimal solutions and select the one that allows him to obtain the higher expected utility in the considered scenario.

\begin{itemize}
\item Existence and calculation of a maximum efficiency incentive compatible direct mechanism.
We wonder if there are some conditions under which the manager can design a mechanism to obtain the same utility it would achieve with compliant users.
The following result provides an answer to this question.
\end{itemize}
\begin{proposition}
If $\forall \, \tau_i \in T_i$ and $\forall \, t_{-i} \in T_{-i}$,
\begin{align}
\left( \dfrac{n + \sum_{j\neq i} t_j + \tau_{i+1}}{n + \sum_{j\neq i} t_j + \tau_i}\right)^{\tau_i + 1} \left( \dfrac{\tau_i}{\tau_{i+1}}\right)^{\tau_i} \geq 1
\label{eq:dis_suf}
\end{align}
then the mechanism $\left(T, D, \overline{d}^S, \overline{\pi} \right) $ where, $\forall \, t \in T$, 
\begin{align}
&\overline{d}^S(t) = d^*(t) \nonumber\\
&\overline{\pi} \left( \cdot \mid t \right) \in \mathcal{F}^{\overline{d}^S,t}
\label{eq:max_eff_flow}
\end{align}
is a maximum efficiency incentive compatible direct mechanism.
\label{prop:max_eff}
\end{proposition}

\begin{IEEEproof}
See Appendix \ref{app:max_eff}
\end{IEEEproof}

\begin{itemize}
\item Characterization of the optimal incentive compatible direct mechanism.
\end{itemize}
In case a maximum efficiency incentive compatible direct mechanism does not exists, 
the manager faces the problem of designing a mechanism 
such that it obtains a utility as close as possible to the maximum efficiency utility.
If $d_0^M \geq \mu$, all additional assumptions made in Subsection \ref{sec:po} are satisfied.  
Hence, according to Proposition \ref{prop}, there exists an optimal incentive compatible direct mechanisms 
such that the intervention device adopts a randomized intervention rule that sustains without 
intervention the suggested action profile.
Such a mechanism can be calculated by solving independently Eqs. (\ref{eq:subprob1}) and (\ref{eq:subprob2}), and (\ref{eq:subprob2}) can be solved considering \emph{pure} affine intervention rules that are simpler to implement than randomized intervention rules.
Moreover, according to Corollary \ref{cor}, the class of affine intervention rules is optimal with respect to $\Gamma$.
Unfortunately, the solution of Eq. (\ref{eq:subprob1}) is hard to compute. 
For this reason, in the following we consider the suboptimal mechanisms proposed in Section \ref{sec:sdm}.

\begin{itemize}
\item Algorithm that converges to an incentive compatible direct mechanism.
\end{itemize}
As for the abstract framework in Subsection \ref{sec:sa}, we consider an algorithm (see Algorithm \ref{algo2}), optimized for the flow control scenario, that converges to an incentive compatible direct mechanism.

\begin{algorithm}
\caption{Flow control algorithm.} \label{algorithm:flow_algo}
\begin{algorithmic}[1]
\STATE \textbf{Initialization}: $\forall \, t \in T$, $d^S(t) = d^*(t)$, $\pi(\tilde{f} \mid t) = 1$ for a certain $\tilde{f} \in \mathcal{F}^{d^S,t}$ and $\pi(f \mid t) = 0$ for $f \neq \tilde{f}$.
\STATE \textbf{For} $s = 1:m$ 
\STATE ~~~\textbf{For} $l = 1:m$
\STATE ~~~~~~\textbf{If} $W_i(\tau_s, \tau_s) < W_i(\tau_s, \tau_l)$ 
\STATE ~~~~~~~~~$d_i^S(\tau_l, t_{-i}) \leftarrow \min\left\lbrace d_i^S(\tau_l, t_{-i}) + \epsilon_i, \; d_i^{NE^0}(\tau_l, t_{-i})  \right\rbrace$, $\pi(\tilde{f} \mid t) \leftarrow 1$ for a certain $\tilde{f} \in \mathcal{F}^{d^S,t}$ and $\pi(f \mid t) = 0$ for $f \neq \tilde{f}$,  $\forall \, t_{-i} \in T_{-i}$ 
\STATE Repeat from $2$ until $4$ is unsatisfied $\forall \, s$ and $l$
\end{algorithmic}
\label{algo2}
\end{algorithm}

\begin{itemize}
\item A priori mechanism.
\end{itemize}

Consider the a priori mechanism where the intervention device, independently of users' types, suggests action profile $\overline{d}$ and adopts the affine intervention rule $\overline{f}$, 
\begin{align}
&\overline{d} = \argmin_d - \ln \left( \mu - \sum_{i=1}^n d_i \right) \mathbb{E}_t \left[ \prod_{i=1}^n d_i^{\frac{t_i}{n}} \right] \nonumber\\
&d_i \geq 0 \;\; ,\;\; d_i \leq \mu \;\; , \;\; \forall i \in \mathcal{N}
\label{eq:geo_opt}
\end{align}

\begin{proposition}
Eq. (\ref{eq:geo_opt}) defines a convex problem if $\tau_m \leq n$. Moreover, if the randomized intervention rule $\pi$ sustains $\overline{d}$ without intervention in $\Gamma$, then $\left( T, D, \overline{d}, \pi \right)$ is an optimal a priori incentive compatible mechanism and the manager's expected utility is $V_0(\overline{d})$.
\label{prop:a_priori}
\end{proposition}

\begin{IEEEproof}
See Appendix \ref{app:a_priori}
\end{IEEEproof}

\subsection{Results}\label{sec:res}

In the following we are going to quantify the manager's expected utility and the expected throughput and delay for each type of user in different scenarios. 
We consider $\mu = 5 \, Mbps$ and a common type set $T_i = \left\lbrace 0.1 , 1\right\rbrace $.
Except for Fig. \ref{fig:1_bis}, we assume that the types are uniformly distributed, i.e., $P(0.1) = P(1) = 0.5$, 
and we plot the results varying the number of users from $2$ to $16$.

\begin{figure}%[htp]
     \centering
          \includegraphics[width=\figw]{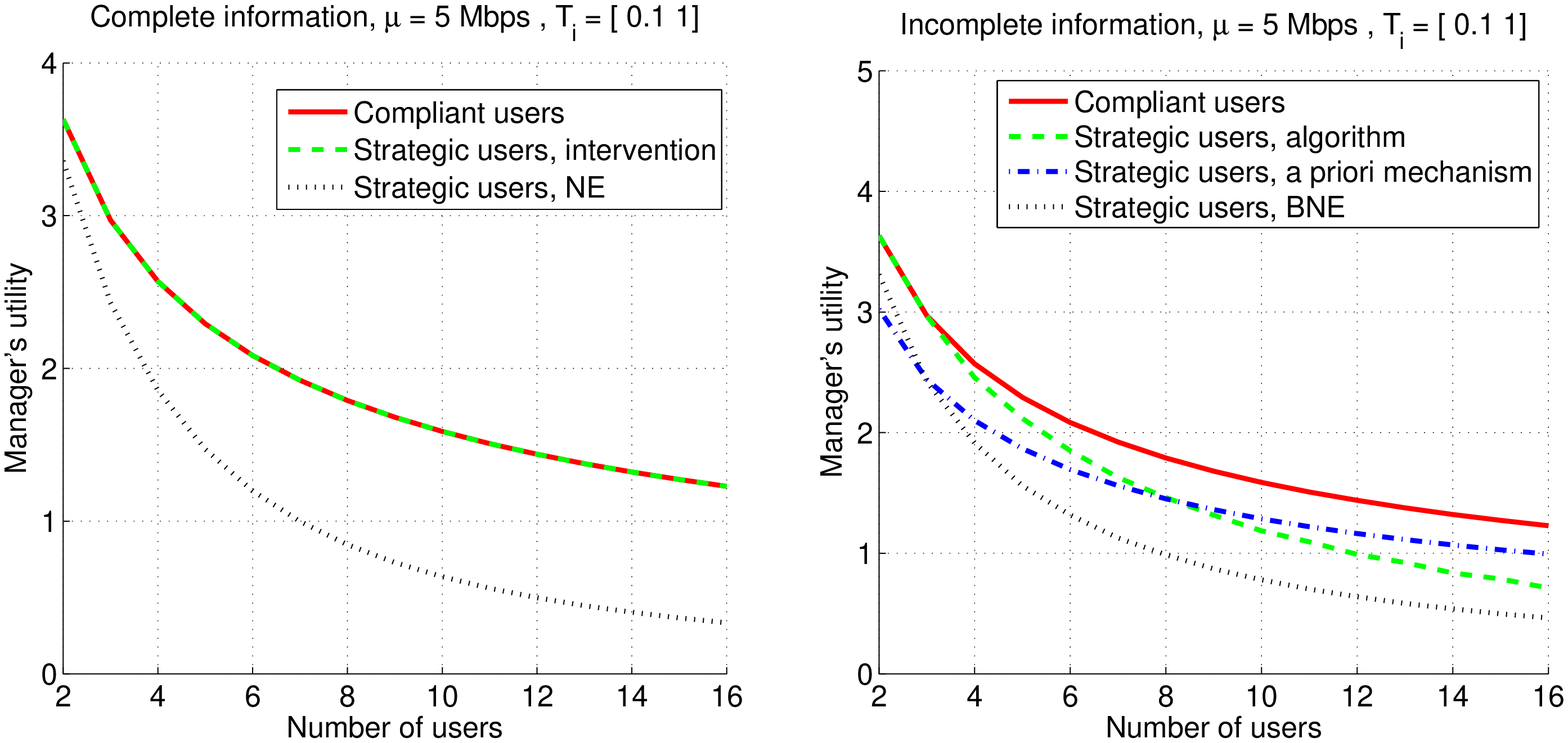}
\caption{Manager's expected utility vs. number of users for the complete and incomplete information scenarios}
\label{fig:1}
\end{figure}

\begin{figure}%[htp]
     \centering
          \includegraphics[width=\figw]{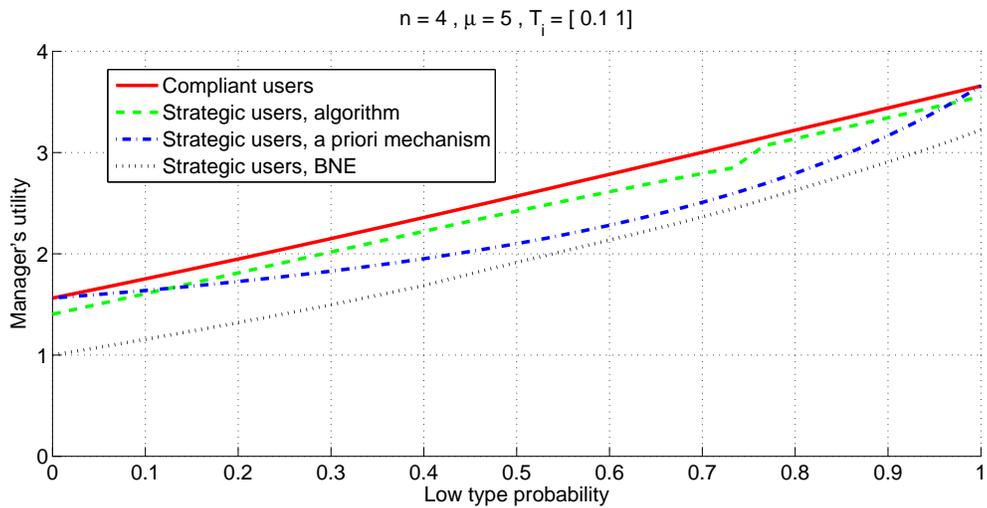}
\caption{Manager's expected utility vs. low type probability for the incomplete information scenario}
\label{fig:1_bis}
\end{figure}

We first look at how the manager's expected utility varies increasing the number of users, in the complete and incomplete
information scenarios. 
The left side of Fig. \ref{fig:1} refers to the complete information scenario. 
The overlapped upper lines represent the manager's expected utility when users are compliant and when they are 
strategic with an intervention device that adopts the optimal intervention rule derived in Subsection \ref{sec:comp_info}. 
The manager's expected utility is decreasing in the number of users because, as the number of users increases, 
the total congestion experienced by every user increases as well. 
However, it is remarkable that with the intervention scheme the manager can completely fill the gap between 
the maximum efficiency utility and its expected utility when the users are strategic
but no incentive scheme is adopted (dotted line).
The right side of Fig. \ref{fig:1} refers to the incomplete information scenario. 
In this scenario the manager is guaranteed to achieve the maximum efficiency utility using the mechanism derived from the algorithm (dashed line) if the number of users is sufficiently small. In fact, for a number of users less than or equal to $3$, it is straightforward to check that the sufficient condition (\ref{eq:dis_suf}) is satisfied, hence, a maximum efficiency mechanism exists and the algorithm converges to it. For a larger number of users, there is no guarantee of optimality, and in fact the results of Fig. 4 show that in this case the manager's expected utility is lower than what could be obtained with compliant users. However, the manager can still considerably increase its expected utility compared to the case of strategic users and no incentive scheme (dotted line), by adopting the mechanism derived from
the algorithm for a number of users lower than $8$ and the a priori mechanism (dash-dot line) for a number of users greater than or equal to $8$ ($\overline{f}$ defined in (\ref{eq:geo_opt_rule}) turns out to sustain the solution of (\ref{eq:geo_opt}) without intervention in $\Gamma$).
It is not surprising that the a priori mechanism is able to obtain good performance for a high number of users, 
in fact in this situation the manager is
able to foresee more accurately the fraction of users of a certain type, hence the information about users'
types becomes less important.

Now we investigate how the results depend on the type probability distribution for the incomplete information 
scenario. In Fig. \ref{fig:1_bis} we fix the number of users to $4$ and we vary the probability of the low type, 
$P(0.1)$, from $0$ to $1$,
which is equivalent to varying $P(1)$ from $1$ to $0$. We can see that the gap between the maximum efficiency utility and the
manager's expected utility achievable with the mechanism derived from the algorithm is not strongly dependent on
the type probability distribution. In fact, such a mechanism provides incentives for each type of user to be honest and
obedient, even though some user types occur rarely.
On the contrary, the a priori mechanism is strongly dependent on the probability distribution of user types. 
In fact, the recommended and enforced action profile depends exclusively on the type probability distribution. 
As an example, if the low type occurs rarely, the intervention device will suggest to the users to adopt an action 
profile that is close to the objective of the users with high type, that will probably be the majority of the 
users in the network. In the extreme case, if low type users are for sure not present in the network
(i.e., P(0.1) = 0), than the adopted action profile will maximize the interests of the users having high type and 
the a priori mechanism is able to achieve the maximum efficiency utility. 
Notice that in this situation the manager has no uncertainty about the types of the users in the network, which is 
the reason why it is able to extract the maximum utility. 
In some sense, the uniform probability distribution represents the worst case for the a priori mechanism because
the manager has the highest uncertainty over the types of the users in the network.

\begin{figure}%[htp]
     \centering
          \includegraphics[width=\figw]{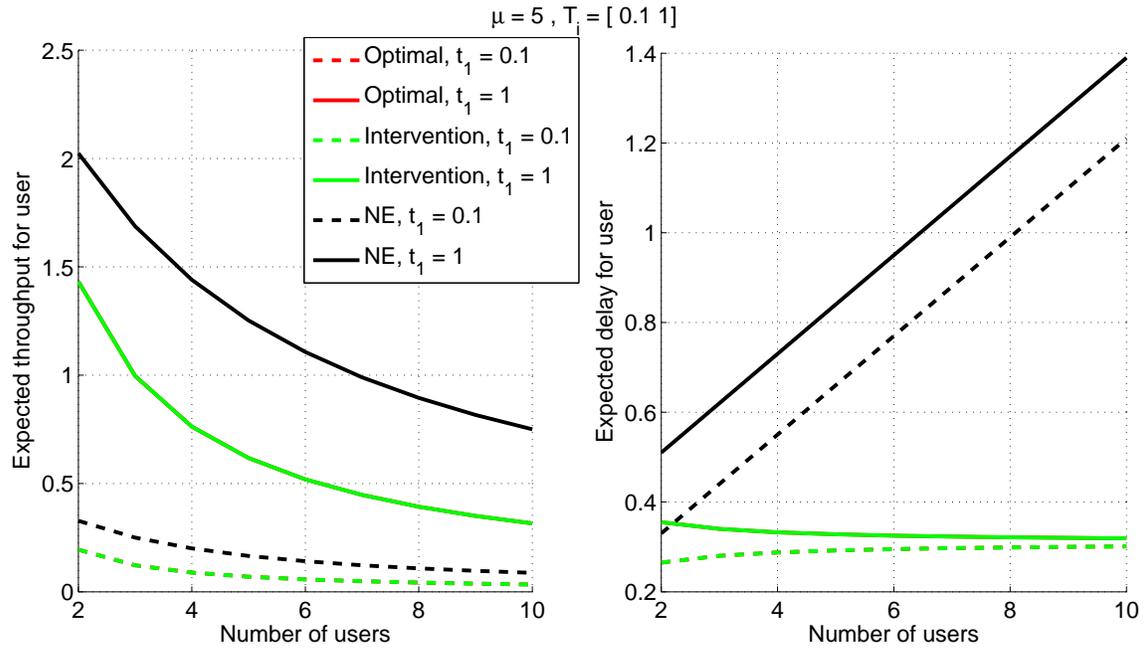}
\caption{Total expected throughput and delay vs. number of users for the complete information scenarios}
\label{fig:2}
\end{figure}

So far we have only considered the utility as performance indicator. 
However, the utility includes the two real performance metrics, the throughput and the delay. 
Now we investigate the expected throughput and delay achievable with the considered schemes in the complete and 
incomplete information scenarios, for each type of user.\footnote{Notice that all users in the network  
experience the same delay. However, such delay depends on the type profile: the higher the number of high type 
users with respect to the number of low type users, the higher the delay.
Thus, the expected delay for a low type user is lower than the expected delay for a high type user.}
Fig. \ref{fig:2} shows the expected throughput (left-side) and 
delay (right-side) for the complete information scenario. 
Continuous lines refer to the high type users, %(i.e., $t_i = 1$), 
while dashed lines refer to the low type users. % (i.e., $t_i = 0.1$).
Notice that the high type users obtain a higher expected throughput and a higher expected delay compared 
to the low type users (this will be true also for the incomplete information scenario), 
confirming that the higher the type the higher the user's preference for throughput with respect to delay. 
In both pictures, the upper (continuous and dashed) lines refer to the strategic scenario without intervention device, in which the users adopt the $NE$ action profile, while the overlapped lower (continuous and dashed) lines represent the optimal action policy, obtainable with compliant users or with strategic users subject to the intervention rule derived in Subsection \ref{sec:comp_info}.
With no incentive scheme, strategic users tend to overuse the resources of the network, transmitting with higher rates compared to the optimal ones.
This translates into much higher delays, that increase quickly as the number of users increases. %, as we can see from the right-side figure.
Conversely, the optimal transmission policy is such that the expected delay is almost constant with respect to the number of users.
This means that also the aggregate throughput is almost constant, and the rate of each user scales as $\frac{1}{n}$.
%In the end, with the optimal policy the users can experience a good throughput keeping a very low delay.

\begin{figure}%[htp]
     \centering
          \includegraphics[width=\figw]{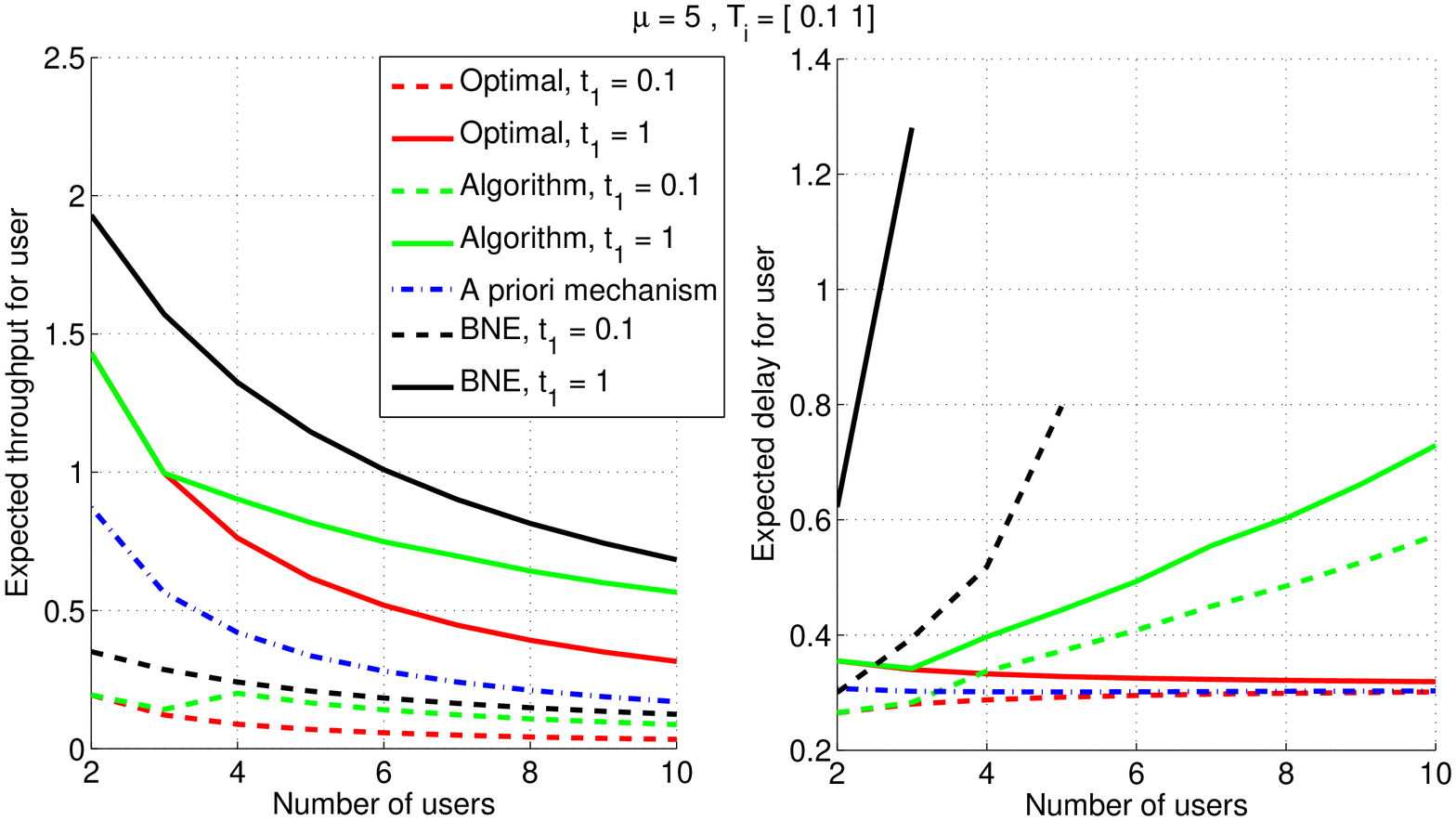}
\caption{Expected throughput and delay for user vs. number of users for the incomplete information scenarios}
\label{fig:3}
\end{figure}

Fig. \ref{fig:3} shows the expected throughput (left-side) and delay (right-side) for the incomplete information scenario. 
Continuous lines refer to the high type users, %(i.e., $t_i = 1$)
while dashed lines refer to the low type users, %(i.e., $t_i = 0.1$) 
with the exception of the performance obtainable adopting the a priori mechanism, represented by the dash-dot line, in which different types of users adopt the same action and experience the same throughput and delay.
In both pictures, the upper (continuous and dashed) lines refer to the strategic scenario without intervention device, in which the users adopt the $BNE$ action profile, while the lower (continuous and dashed) lines represent the optimal action policy.
The performance obtainable adopting the mechanism derived from the algorithm lies between them.
%and it is equal to the optimal performance for a number of user lower than or equal to $3$.
The lines that represent the expected delay for the $BNE$ action profile are truncated for a number of users 
equal to $3$ and $5$ because the system might become unstable.
In fact, in the $BNE$ the expected utility of a user is maximized, given that the other users adopt the $BNE$.
However, for some type profile instances, the utility might be equal to $0$, i.e., the delay might diverge. 
Thus, the expected delay diverges as well.
In words, there is a positive probability that the network becomes congested.
The mechanism derived from the algorithm allows to improve this situation, 
limiting the delay experienced by each user.
However, such a delay increases almost linearly as the number of users increases.
This is the reason why the a priori mechanism, at a certain point, even though it is not able to differentiate the service given to different classes of traffic, is able to obtain a better performance (from the manager's utility point of view) than the mechanism derived from the algorithm, 
In the a priori mechanism each user, independently of its type, adopts a rate which is between the optimal 
rates adopted by the low type users and the high type users, and this situation reflects in the expected delay.
This allows to keep a very low and constant delay with respect to the number of users.

% OLD FIGURES
%\begin{figure}%[htp]
%     \centering
%          \includegraphics[width=\figw]{UtilityVsUs_n16.eps}
%\caption{$.$}
%\label{fig:n4_m4}
%\end{figure}
%
%\begin{figure}%[htp]
%     \centering
%          \includegraphics[width=\figw]{ThroughVsUs_n16.eps}
%\caption{$.$}
%\label{fig:n4_m4}
%\end{figure}
%
%\begin{figure}%[htp]
%     \centering
%          \includegraphics[width=\figw]{DelayVsUs_n16.eps}
%\caption{$.$}
%\label{fig:n4_m4}
%\end{figure}
%
%
%\begin{figure}%[htp]
%     \centering
%          \includegraphics[width=\figw]{ThTypeVsUs_n16.eps}
%\caption{$.$}
%\label{fig:n4_m4}
%\end{figure}
%
%
%
%\begin{figure}%[htp]
%     \centering
%          \includegraphics[width=\figw]{DeTypeVsUs_n16.eps}
%\caption{$.$}
%\label{fig:n4_m4}
%\end{figure}

\section{Conclusion}\label{sec:conc}

In this paper we extend the intervention framework introduced by \cite{ParkMihaela_JSAC} to take into account 
situations in which users hold relevant information that the manager cannot observe.
To design a system that is efficient and robust to self-interested strategic users, the manager must provide the incentives for the users to report truthfully and to follow the recommendations. 
For a class of environments that includes many resource allocation games in communication networks, we provide 
conditions under which it is possible for the manager to achieve its benchmark optimum and conditions under which it is impossible for the manager to achieve its benchmark optimum. 
In both cases, we are able to characterize the optimal coordination mechanism the manager should adopt.
Although we can characterize the optimal mechanism, we also describe a suboptimal mechanism that is easy 
to compute and a suboptimal mechanism that does not rely on the communication between the users and the intervention device.
Finally, we apply our framework and results to the design of a flow control management system.
Computations show that the considered schemes can considerably improve the manager's utility.

\appendices

\section{Proof of Proposition \ref{lemma1}}\label{app:contradiction}

\begin{IEEEproof}

$\Rightarrow$

We prove the result by contradiction.

$\left( T, D, d^S, \pi \right) $ is a maximum efficiency incentive compatible direct mechanism.
Suppose that $\exists \, \hat{t}$ such that $d^S(\hat{t}) \neq d^*(\hat{t})$, then
\begin{align}
& \argmax_{d^S, \pi} V_0 \left( \phi^*, \delta^* \right) = \argmax_{d^S, \pi} \sum_{t \in T} \sum_{f \in \mathcal{F}} P_t(t) \pi \left( f \mid t \right) U_0 \left( f, d^S(t), t \right) < \nonumber\\ 
&< \argmax_{d^S, \pi} \sum_{t \in T, t \neq \hat{t}} \sum_{f \in \mathcal{F}} P_t(t) \pi \left( f \mid t \right) U_0 \left( f, d^S(t), t \right) + P_t(\hat{t}) U_0(d_0^*, d^*(\hat{t}), \hat{t}) \leq V_0^{ME}
\label{eq:pippo}
\end{align}

Now suppose that $\exists \, \hat{t}$ such that $\pi \left( \cdot \mid \hat{t} \right) \notin \mathcal{F}^{d^*(\hat{t}),\hat{t}}$.
If $\pi \left( \cdot \mid \hat{t} \right)$ sustains $d^*(\hat{t})$ but $\exists \, \hat{f}$ such that $\pi \left( \hat{f} \mid \hat{t} \right) > 0$ and $\hat{f} \left( d^*(\hat{t}) \right) \neq d_0^*$, then
\begin{align}
& \argmax_{d^s, \pi} V_0 \left( \phi^*, \delta^* \right) = \argmax_{d^s, \pi} \sum_{t \in T} \sum_{f \in \mathcal{F}} P_t(t) \pi \left( f \mid t \right) U_0 \left( f, d^S, t \right) \leq \nonumber\\ 
&\leq \sum_{t \in T, t \neq \hat{t}} P_t(t) U_0(d_0^*, d^*(t), t) + (1 - \pi(\hat{f} \mid \hat{t})) U_0(d_0^*, d^*(\hat{t}), \hat{t}) + \pi(\hat{f} \mid \hat{t}) U_0(\hat{f}, d^S(\hat{t}), \hat{t}) < \nonumber\\
&< \sum_{t \in T, t \neq \hat{t}} P_t(t) U_0(d_0^*, d^*(t), t) + (1 - \pi(\hat{f} \mid \hat{t})) U_0(d_0^*, d^*(\hat{t}), \hat{t}) + \pi(\hat{f} \mid \hat{t}) U_0(d_0^*, d^*(\hat{t}), \hat{t}) = V_0^{ME}
\label{eq:pippo}
\end{align}
If $\pi \left( \cdot \mid \hat{t} \right)$ does not sustain $d^*(\hat{t})$, then $\exists \, i$ and $\hat{d}_i$ such that $\overline{U}_i \left( d^*(\hat{t}), t \right) < \overline{U}_i \left( \hat{d}_i, d^*_{-i}(\hat{t}), t \right)$.
In this case the intervention device is not able to provide incentive to user $i$ to adopt optimal strategy  $d^*_{i}(\hat{t})$ when the type profile is $\hat{t}$, therefore the mechanism is not incentive compatible.

Finally, \textbf{2} is a particular case of the incentive-compatibility constraints of \textbf{OICDM}, therefore it must be satisfied.

$\Leftarrow$

It is straightforward to verify that a mechanism satisfying $\textbf{1}-\textbf{4}$ is incentive compatible and the utility of the intervention device is equal to Eq. (\ref{eq:v0_ME}).

\end{IEEEproof}

\section{Proof of Lemma \ref{prop:kakutani}}\label{app:kakutani}

\begin{IEEEproof}

Let $\left( T, D, d^S, \pi \right)$ be an optimal incentive compatible direct mechanism.

Given a type profile $t$, we use the notations
\begin{align}
&D^S_i = \left[ d_i^{min}, \; \min \left\lbrace d_i^S(t)\, , \, d_i^{NE^0}(t) \right\rbrace \right] \;\; , \;\; D^S = D^S_1 \times \cdots \times D^S_n \;\; , \;\; D^S_{-i} = D^S \setminus D^S_i \nonumber\\
&a_i(t) = \mathbb{E}_f \left[ U_i \left( f, d^S(t), t \right) \right]
\end{align} 

We define the function $g_i(d_{-i})$ in the domain $D^S_{-i}$ as follows:
\begin{align}
g_i(d_{-i}) = \left\lbrace d_i \in D^S_i \; \mbox{such that} \; U_i \left( d_0^*, d, t \right) = a_i \right\rbrace 
\end{align} 
The function $g_i$ is a non-empty set-valued function from $D^S_{-i}$ to the power set of $D^S_{-i}$.
In fact, $\forall \, t \in T$ and $d_{-i} \in D^S_{-i}$,
\begin{equation}
U_i \left( d_0^*, d_i^{min}, d_{-i}, t \right)  = 0 \leq a_i \leq U_i \left( d_0^*, d^S_i(t), d^S_{-i}(t), t \right) \leq  U_i \left( d_0^*, d^S_i(t), d_{-i}, t \right)
\label{last}
\end{equation}
The second inequality of Eq. (\ref{last}) is valid because $i$'s utility is non increasing with respect to the intervention 
level, i.e., $U_i \left( f, d^S(t), t \right) \leq U_i \left( d_0^*, d^S(t), t \right) $, $\forall \, f$, 
which implies that $\mathbb{E}_f \left[ U_i \left( f, d^S(t), t \right) \right] \leq U_i \left( d_0^*, d^S(t), t \right)$.
The last inequality of Eq. (\ref{last}) is valid because $i$'s utility is non increasing in the actions of the other users
and, from the definition of the set $D^S_i$, $d^S_{-i}(t) \geq d_{-i}$, $\forall \, d_{-i} \in D^S_i$.
Eq. (\ref{last}) and the continuity of $i$'s utility imply that an action $\hat{d}_i \in D^S_i$ satisfying 
$U_i \left( d_0^*, \hat{d}_i, d_{-i}, t \right) = a_i$ exists, $\forall d_{-i} \in D^S_{-i}$.
Moreover, by definition $g_i$ has a closed graph (i.e., the graph of $g_i$ is a closed subset of $D^S_{-i} \times D^S_i$) and, since $i$'s utility is non decreasing in $\left[ d_i^{min} \; d_i^{NE^0}(t) \right]$, $g_i(d_{-i})$ is convex, $\forall \, d_{-i} \in D^S_{-i}$.

We define the function $g(d) = \left( g_1(d_{-1}), \cdots, g_n(d_{-n}) \right) $, $\forall \, d \in D^S$.
$g$ is defined from the non-empty, compact and convex set $D^S$ to the power set of $D^S$.
Thanks to the properties of $g_i$, $g$ has a closed graph and $g(d)$ is non-empty and convex.
Therefore we can apply Kakutani fixed-point theorem \cite{Kakutani} to affirm that a fixed point exists, i.e., there exists an 
action profile $\hat{d} \in D^S$ such that $U_i \left( d_0^*, \hat{d}, t \right) = a_i$, $\forall \, i \in \mathcal{N}$.
For each type profile $t \in T$ there exists a different fixed point, hence, we use the notation $\hat{d}(t)$.
Notice that $\hat{d}(t) < d^{NE^0}(t)$, therefore the intervention device is able to sustain $\hat{d}(t)$ without intervention.

Finally, the original optimal mechanism can be substituted by a mechanism where, $\forall \, t \in T$, 
the intervention device suggests $\hat{d}(t)$ and adopts a randomized intervention rule able 
to sustain it without intervention. 
In the new mechanism, the users are obedient because the intervention rule sustains $\hat{d}(t)$ and they 
are honest because the utilities they obtain for each combination of reports are the same as in the 
original incentive compatible mechanism.
The utility of the intervention device, which depends only on the users' utilities, is the same as in the 
original mechanism.
Therefore we have obtained an optimal incentive compatible direct mechanism where the intervention device 
adopts a randomized intervention rule that sustains without intervention the suggested action profile.

\end{IEEEproof}

\section{Proof of Proposition \ref{prop:BNE}}\label{app:matrix_inv}

\begin{IEEEproof}

\begin{align}
V_i(d, t_i) &= \mathbb{E}_{t_{-i}} \left[ U_i(d, t_i) \right] =  d_i(t_i)^{t_i} \mathbb{E}_{t_{-i}} \left[ \left( \mu - \lambda \right) \right] =  d_i(t_i)^{t_i}  \left[ \left( \mu - d_i(t_i) - \sum_{j=1, j\neq i}^n  \mathbb{E}_{t_j} \left[ d_j(t_j) \right] \right) \right]  \nonumber\\
\dfrac{\partial \ln V_i(d, t_i)}{\partial d_i(t_i)} &= \dfrac{t_i}{d_i(t_i)} - \dfrac{1}{\mu - d_i(t_i) - \sum_{j=1, j\neq i}^n  \mathbb{E}_{t_j} \left[ d_j(t_j) \right]} \nonumber\\
\dfrac{\partial^2 \ln V_i(d, t_i)}{\partial d_i^2(t_i)} &= - \dfrac{t_i}{d_i^2(t_i)} - \dfrac{1}{\left( \mu - d_i(t_i) - \sum_{j=1, j\neq i}^n  \mathbb{E}_{t_j} \left[ d_j(t_j) \right] \right)^2 } < 0
\label{eq:V_cos}
\end{align} 
 
Imposing that the first derivative is equal to $0$, we obtain that the Bayesian Nash Equilibrium $d^{BNE}$ must satisfy, $\forall \, i \in \mathcal{N}$ and $ \forall \, l = 1, \dots, m$, 
\begin{align}
\left( 1 + \tau_l \right) d_i^{BNE}(\tau_l) + \tau_l \sum_{j=1, j\neq i}^n \sum_{k=1}^m P(\tau_k) d_j^{BNE}(\tau_k) = \mu \tau_l 
\label{eq:deri}
\end{align} 
%assuming that the values obtained in this way are such that $d_i^{\tau_k} \leq \overline{d_i}$, $\forall \tau_k \in T_i$ and $i \in \mathcal{N}$.

The system of equations defined by (\ref{eq:deri}) can be written as a matrix equation of the form 
\begin{align}
\mathbf{A} d^{BNE} = b
\label{eq:system}
\end{align}
where 
\begin{equation}
d^{BNE} = \left[\begin{array}{c} d_1^{BNE} \\ \vdots \\ d_n^{BNE} \end{array}\right], \;\;\;
d_i^{BNE} = \left[\begin{array}{c} d_i^{BNE}(\tau_1) \\ \vdots \\ d_i^{BNE}(\tau_m) \end{array}\right], \;\;\;
b = \left[\begin{array}{c} \hat{b} \\ \vdots \\ \hat{b} \end{array}\right], \;\;\;
\hat{b} = \left[\begin{array}{c} \mu \tau_1 \\ \vdots \\  \mu \tau_m \end{array}\right],
\label{eq:matrix}
\end{equation}
\begin{equation}
\mathbf{A} = \left[\begin{array}{cccc} \mathbf{\Lambda} & \mathbf{\tau}\cdot\mathbf{P} & \cdots & \mathbf{\tau}\cdot\mathbf{P} \\
\mathbf{\tau}\cdot\mathbf{P} & \mathbf{\Lambda} & \cdots & \mathbf{\tau}\cdot\mathbf{P} \\ \vdots & \vdots & \ddots & \vdots \\
\mathbf{\tau}\cdot\mathbf{P} & \mathbf{\tau}\cdot\mathbf{P} & \cdots & \mathbf{\Lambda} \end{array}\right], 
\label{eq:matrix}
\end{equation}
\begin{equation}
\mathbf{\Lambda} = \mathrm{diag}\left(1+\tau_1,\ldots,1+\tau_m\right), \;\;\;
\tau = \left[\begin{array}{c} \tau_1 \\ \vdots \\ \tau_m \end{array}\right], \;\;\; 
%x = \left[\begin{array}{c} d_1(\tau_1) \\ \vdots \\ d_1(\tau_m) \\ d_2(\tau_1) \\ \vdots \\ d_2(\tau_m) \\ \vdots \\ d_n(\tau_m) \end{array}\right], \;\;\;
\mathbf{P} = \left[\begin{array}{ccc} P(\tau_1) & \ldots & P(\tau_m) \end{array}\right].
\end{equation}

We want to compute the inverse of the matrix $\mathbf{A}$.
We can write $\mathbf{A}$ as
\begin{equation}
\mathbf{A} = \left[\begin{array}{ccc} \mathbf{\Lambda}-\tau\cdot\mathbf{P} & & \\
 & \ddots & \\ & & \mathbf{\Lambda}-\tau\cdot\mathbf{P} \end{array}\right] + \left[\begin{array}{c} \mathbf{I} \\ \vdots \\ \mathbf{I} \end{array}\right]\cdot\left[\begin{array}{ccc} \tau\cdot\mathbf{P} & \ldots & \tau\cdot\mathbf{P} \end{array}\right]
\end{equation}
where $\mathbf{I}$ is the identity matrix in $\mathbb{R}^{m\times m}$. 

The matrix inversion Lemma states that
\begin{equation}
(\mathbf{E}+\mathbf{BCD} )^{-1} = \mathbf{E}^{-1}-\mathbf{E}^{-1}\mathbf{B}\left(\mathbf{C}^{-1}+\mathbf{D} \mathbf{E}^{-1} \mathbf{B}\right)^{-1} \mathbf{D} \mathbf{E}^{-1}
\end{equation}

Applying the matrix inversion Lemma to $\mathbf{A}^{-1}$ we obtain
\begin{eqnarray}
\mathbf{A}^{-1} = 
\left[\begin{array}{ccc} \mathbf{\Lambda}-\tau\cdot\mathbf{P} & & \\ & \ddots & \\ & & \mathbf{\Lambda}-\tau\cdot\mathbf{P}
\end{array}\right]^{-1} 
- \left[\begin{array}{ccc} \mathbf{\Lambda}-\tau\cdot\mathbf{P} & & \\ & \ddots & \\ & & \mathbf{\Lambda}-\tau\cdot\mathbf{P} \end{array}\right]^{-1} 
\cdot \left[\begin{array}{c} \mathbf{I} \\ \vdots \\ \mathbf{I} \end{array}\right] \cdot \nonumber\\
\left(\underbrace{\mathbf{I}^{-1}+\left[\begin{array}{ccc} \tau\cdot\mathbf{P} & \ldots & \tau\cdot\mathbf{P} \end{array}\right] \cdot
\left[\begin{array}{ccc}
\mathbf{\Lambda}-\tau\cdot\mathbf{P} & & \\ & \ddots & \\ & & \mathbf{\Lambda}-\tau\cdot\mathbf{P} \end{array}\right]^{-1}\left[\begin{array}{c} \mathbf{I} \nonumber\\ \vdots \\ \mathbf{I} \end{array}\right]}_{\mathbf{Y}}\right)^{-1} \cdot \nonumber\\
\left[\begin{array}{ccc} \tau\cdot\mathbf{P} & \ldots & \tau\cdot\mathbf{P} \end{array}\right] \cdot \left[\begin{array}{ccc}
\mathbf{\Lambda}-\tau\cdot\mathbf{P} & & \\ & \ddots & \\ & & \mathbf{\Lambda}-\tau\cdot\mathbf{P} \end{array}\right]^{-1}
\end{eqnarray}

First, we calculate
\begin{eqnarray}
\left(\mathbf{\Lambda}-\tau\cdot\mathbf{P}\right)^{-1} & = & \mathbf{\Lambda}^{-1}-\mathbf{\Lambda}^{-1}\cdot \tau \cdot
\left(-1+\mathbf{P}\cdot\mathbf{\Lambda}^{-1}\cdot\tau\right)^{-1}\cdot\mathbf{P}\cdot\mathbf{\Lambda}^{-1} \nonumber\\
& = & \mathbf{\Lambda}^{-1}-\mathbf{\Lambda}^{-1}\cdot \tau \cdot \frac{1}{-1+\sum_{i=1}^m
P(\tau_i)\frac{\tau_i}{1+\tau_i}} \cdot\mathbf{P}\cdot\mathbf{\Lambda}^{-1} \nonumber\\
& = & \mathbf{\Lambda}^{-1}-\mathbf{\Lambda}^{-1}\cdot \tau \cdot \beta \cdot\mathbf{P}\cdot\mathbf{\Lambda}^{-1}
\end{eqnarray}
where $\beta = \frac{1}{-1+\sum_{i=1}^m P(\tau_i)\frac{\tau_i}{1+\tau_i}}$.

Now we calculate $\mathbf{Y}^{-1}$. 
We rewrite $\mathbf{Y}$ as
\begin{eqnarray}
\mathbf{Y} & = & \mathbf{I}+\left[\begin{array}{ccc} \tau\cdot\mathbf{P} & \ldots & \tau\cdot\mathbf{P}
\end{array}\right] \cdot \left[\begin{array}{ccc} \mathbf{\Lambda}-\tau\cdot\mathbf{P} & & \\ & \ddots & \\ & & \mathbf{\Lambda}-\tau\cdot\mathbf{P}
\end{array}\right]^{-1}\left[\begin{array}{c} \mathbf{I} \\ \vdots \\ \mathbf{I} \end{array}\right] \nonumber\\
& = & \mathbf{I}+\left[\begin{array}{ccc} \tau\cdot\mathbf{P} & \ldots & \tau\cdot\mathbf{P}
\end{array}\right] \cdot \left[\begin{array}{ccc} \mathbf{\Lambda}^{-1}-\beta \mathbf{\Lambda}^{-1} \tau \mathbf{P} \mathbf{\Lambda}^{-1} & & \\ & \ddots & \\ & & \mathbf{\Lambda}^{-1}-\beta \mathbf{\Lambda}^{-1} \tau \mathbf{P} \mathbf{\Lambda}^{-1}
\end{array}\right] \cdot \left[\begin{array}{c} \mathbf{I} \\ \vdots \\ \mathbf{I} \end{array}\right] \nonumber\\
& = & \mathbf{I}+n\cdot \tau\cdot\mathbf{P} \cdot \left(\mathbf{\Lambda}^{-1}-\beta \mathbf{\Lambda}^{-1} \tau \mathbf{P}
\mathbf{\Lambda}^{-1}\right) 
%& = & \underbrace{\mathbf{I}}_{A} + \underbrace{\tau}_{B}\cdot
%\underbrace{\left[n\cdot\left(1-\mathbf{P}\mathbf{\Lambda}^{-1}\tau\beta\right)\right]}_{C} \cdot \underbrace{\mathbf{P} \mathbf{\Lambda}^{-1}}_{D}
 =  \mathbf{I} + \tau \cdot \left[n\cdot\left(1-\mathbf{P}\mathbf{\Lambda}^{-1}\tau\beta\right)\right] \cdot \mathbf{P} \mathbf{\Lambda}^{-1} \nonumber\\
& = & \mathbf{I} + \tau \cdot \frac{n}{1-\sum_{i=1}^m P(\tau_i) \frac{\tau_i}{1+\tau_i}} \cdot \mathbf{P} \mathbf{\Lambda}^{-1}
\end{eqnarray}

Applying the matrix inversion Lemma to $\mathbf{Y}^{-1}$ we obtain 
\begin{eqnarray}
\mathbf{Y}^{-1} & = & \mathbf{I}^{-1} - \mathbf{I}^{-1} \tau \cdot \left( \frac{1-\sum_{i=1}^m P(\tau_i) \frac{\tau_i}{1+\tau_i}}{n} + \mathbf{P} \cdot \mathbf{\Lambda}^{-1} \cdot \mathbf{I}^{-1} \cdot \tau \right)^{-1} \cdot \mathbf{P} \cdot \mathbf{\Lambda}^{-1} \cdot \mathbf{I}^{-1} \nonumber\\ 
& = & \mathbf{I} - \tau \cdot \left( \frac{1}{\frac{1-\sum_{i=1}^m P(\tau_i) \frac{\tau_i}{1+\tau_i}}{n}+\sum_{i=1}^m P(\tau_i) \frac{\tau_i}{1+\tau_i}} \right) \cdot \mathbf{P} \cdot \mathbf{\Lambda}^{-1} \nonumber\\
& = & \mathbf{I} -  \frac{n}{1+(n-1)\sum_{i=1}^m P(\tau_i) \frac{\tau_i}{1+\tau_i}} \cdot \tau \mathbf{P} \mathbf{\Lambda}^{-1}
\end{eqnarray}

Finally, we can calculate $\mathbf{A}^{-1}$ as
\begin{eqnarray}
\mathbf{A}^{-1} & = & \left[\begin{array}{ccc} \mathbf{\Lambda}-\tau\cdot\mathbf{P} & & \\ & \ddots & \\ & & \mathbf{\Lambda}-\tau\cdot\mathbf{P}
\end{array}\right]^{-1} - \left[\begin{array}{ccc} \mathbf{\Lambda}-\tau\cdot\mathbf{P} & & \\ & \ddots & \\ & & \mathbf{\Lambda}-\tau\cdot\mathbf{P} \end{array}\right]^{-1} \cdot \left[\begin{array}{c} \mathbf{I} \\ \vdots \\ \mathbf{I} \end{array}\right] \cdot \nonumber\\
& = & \left(\mathbf{I} -  \frac{n}{1+(n-1)\sum_{i=1}^m P(\tau_i) \frac{\tau_i}{1+\tau_i}} \cdot \tau \mathbf{P} \mathbf{\Lambda}^{-1}\right) \cdot  
\left[\begin{array}{ccc} \tau\cdot\mathbf{P} & \ldots & \tau\cdot\mathbf{P} \end{array}\right] \cdot \nonumber\\ && \left[\begin{array}{ccc}
\mathbf{\Lambda}-\tau\cdot\mathbf{P} & & \\ & \ddots & \\ & & \mathbf{\Lambda}-\tau\cdot\mathbf{P} \end{array}\right]^{-1} =  \left[\begin{array}{ccc} \mathbf{B} & & \\ & \ddots & \\ & & \mathbf{B}
\end{array}\right] - \left[\begin{array}{ccc} \mathbf{C} & \ldots & \mathbf{C} \\ \vdots & \ddots & \vdots \\ \mathbf{C} & \ldots & \mathbf{C}
\end{array}\right]
\end{eqnarray}
where
\begin{eqnarray}
\mathbf{B} &=& \mathbf{\Lambda}^{-1} - \beta \mathbf{\Lambda}^{-1} \tau \mathbf{P} \mathbf{\Lambda}^{-1} \nonumber\\
\mathbf{C} &=& \left(\mathbf{\Lambda}^{-1} - \beta \mathbf{\Lambda}^{-1} \tau \mathbf{P} \mathbf{\Lambda}^{-1}\right)\cdot\left(\mathbf{I} -
\frac{n}{1+(n-1)\sum_{i=1}^m P(\tau_i) \frac{\tau_i}{1+\tau_i}} \cdot \tau \mathbf{P} \mathbf{\Lambda}^{-1}\right)\cdot \tau \mathbf{P} \cdot
\left(\mathbf{\Lambda}^{-1} - \beta \mathbf{\Lambda}^{-1} \tau \mathbf{P} \mathbf{\Lambda}^{-1}\right) \nonumber
\end{eqnarray}

Hence, the $BNE$ can be analytically computed:
\begin{align}
d^{BNE} = A^{-1} b
\end{align}

\end{IEEEproof}

\section{Proof of Lemma \ref{lemma_interv}}\label{app:interv_NE}

\begin{IEEEproof}

We study the sign of the derivative of $i$'s utility with respect to $i$' action
\begin{equation}
\dfrac{\partial \ln U_i(f, d_i, \tilde{d}_{-i}, t)}{\partial d_i} =
\left\lbrace
\begin{array}{lll}
\dfrac{t_i}{d_i} - \dfrac{1}{\mu - \sum_{k\neq i} \tilde{d}_k - d_i} & d_i < \tilde{d}_i \\
\dfrac{t_i}{d_i} - \dfrac{1+c_i}{\mu - \sum_{k\neq i} \tilde{d}_k - d_i - c_i (d_i - \tilde{d}_i)} & \tilde{d}_i < d_i < \tilde{d}_i + \dfrac{d_0^M}{c_i} \\
\dfrac{t_i}{d_i} - \dfrac{1}{\mu - \sum_{k\neq i} \tilde{d}_k - d_i - d_0^M} & d_i > \tilde{d}_i + \dfrac{d_0^M}{c_i}
\end{array}
\right. \nonumber
\label{eq:partial}
\end{equation}

We denote by $d_i^{BR}(d_{-i})$ the best response function of user $i$, i.e., $i$'s action that maximizes $i$'s utility when the action vector of the other users is $d_{-i}$.
Since the users' utilities satisfy the assumptions \textbf{A4}-\textbf{A6}, $\dfrac{\partial U_i(f, d_i, \tilde{d}_{-i}, t)}{\partial d_i} \geq 0$ for $d_i < \tilde{d}_i$.
In fact $U_i(f, d, t_i)$ is increasing with respect to $d_i$ in $\left[ 0, \; d_i^{BR}(\tilde{d}_{-i}) \right) $ and $\tilde{d}_i \leq d_i^{NE^0} = d_i^{BR}(d_{-i}^{NE^0}) \leq d_i^{BR}(\tilde{d}_{-i})$, 
where the first inequality is an hypothesis of the Lemma and the last inequality is valid because of 
the submodularity of the game.

Imposing the condition $\dfrac{\partial U_i(f, d_i, \tilde{d}_{-i}, t)}{\partial d_i} \leq 0$ in $\tilde{d}_i < d_i < \tilde{d}_i + \dfrac{d_0^M}{c_i}$, we find 
\begin{equation}
c_i \geq \dfrac{t_i \left( \mu - \sum_{k=1, k\neq i}^n \tilde{d}_k - d_i \right) - d_i}{t_i \left( d_i - \tilde{d}_i \right) + d_i}
\label{disney}
\end{equation}
The right hand side term of \ref{disney} is decreasing in $d_i$, therefore the condition is valid in $\tilde{d}_i < d_i < \tilde{d}_i + \dfrac{d_0^M}{c_i}$ 
if and only if it is valid in $\tilde{d}_i$, obtaining 
% the condition of Eq. (\ref{eq:conditions}).
\begin{equation}
c_i \geq \dfrac{t_i \left( \mu - \sum_{k=1}^n \tilde{d}_k \right) - \tilde{d}_i}{\tilde{d}_i}
\label{eq}
\end{equation}
Notice that the condition on $c_i$ is a necessary condition for $\tilde{d}_i$ to be a $NE$.
In fact if it is not satisfied then $U_i(f, d_i, \tilde{d}_{-i}, t)$ is strictly increasing in $\tilde{d}_i$ and, for the continuity of $U_i(f, d_i, \tilde{d}_{-i}, t)$ with respect to $d_i$, we can find an action $\hat{d}_i > \tilde{d}_i$ such that $U_i(f, \hat{d}_i, \tilde{d}_{-i}, t) > U_i(f, \tilde{d}_i, \tilde{d}_{-i}, t)$.

Finally, imposing the condition $\dfrac{\partial U_i(f, d_i, \tilde{d}_{-i}, t)}{\partial d_i} \leq 0$ in $d_i > \tilde{d}_i + \dfrac{d_0^M}{c_i}$, we find
\begin{equation}
d_0^M \geq \dfrac{c_i \left[ t_i \left( \mu - \sum_{k=1}^n \tilde{d}_k \right) - \tilde{d}_i \right] }{1 + t_i (1 + c_i)} 
\end{equation}
Notice that, given the condition on $c_i$, this last condition is sufficient for $\tilde{d}_i$ to be a global maximizer. 
In fact in this way $U_i(f, d_i, \tilde{d}_{-i}, t)$ becomes quasi-concave in $d_i$: increasing for $d_i <  \tilde{d}_i$ and decreasing for $d_i >  \tilde{d}_i$.

\end{IEEEproof}

\section{Proof of Proposition \ref{prop:max_eff}}\label{app:max_eff}

\begin{IEEEproof}

Condition \textbf{1}, \textbf{3} and \textbf{4} of \ref{lemma1} are satisfied (we implicitly assume that $d_0^M \geq \dfrac{\mu}{1+\tau_1}$). 
It remains to verify that \textbf{2} is satisfied, i.e., $\forall \tau_i, \hat{\tau}_i \in T_i$, 
\begin{align}
\sum_{t_{-i} \in T_{-i}} P(t_{-i}) d_i^{\tau_i} \left( \mu - \sum_{k=1}^n d_k \right) \geq  \sum_{t_{-i} \in T_{-i}} P(t_{-i}) \hat{d}_i^{\tau_i} \left( \mu - \sum_{k=1}^n \hat{d}_k \right)
\label{eq:100}
\end{align}
where, $\forall j \neq i$,
\begin{align}
d_i = \dfrac{\tau_i \mu}{n + \sum_{k \neq i} t_k + \tau_i } \;\; , \;\; 
d_j = \dfrac{t_j \mu}{n + \sum_{k \neq i} t_k + \tau_i } \;\; , \;\; 
\hat{d}_i = \dfrac{\hat{\tau}_i \mu}{n + \sum_{k \neq i} t_k + \hat{\tau}_i } \;\; , \;\; 
\hat{d}_j = \dfrac{t_j \mu}{n + \sum_{k \neq i} t_k + \hat{\tau}_i } 
\label{eq:101}
\end{align}

In particular, Eq. (\ref{eq:100}) is valid if, $\forall t_{-i} \in T_{-i}$,
\begin{align}
d_i^{\tau_i} \left( \mu - \sum_{k=1}^n d_k \right) \geq  \hat{d}_i^{\tau_i} \left( \mu - \sum_{k=1}^n \hat{d}_k \right)
\label{eq:102}
\end{align}

Substituting Eq. (\ref{eq:101}) into Eq. (\ref{eq:102}) we obtain:
\begin{align}
\left( \dfrac{n + \sum_{k \neq i} t_k + \hat{\tau}_i}{n + \sum_{k \neq i} t_k + \tau_i} \right)^{\tau_i+1}     
\left( \dfrac{\tau_i}{\hat{\tau}_i} \right)^{\tau_i} \geq 1
\label{eq:103}
\end{align}

We use the notation $a = n + \sum_{k \neq i} t_k$, and $x = \frac{\hat{\tau}_i}{\tau_i}$.
We want to find the condition on $\tau_i$ and $x$ such that
\begin{align}
f(x) = \left( \dfrac{a + \tau_i x}{a + \tau_i} \right)^{\tau_i+1} x^{-\tau_i} \geq 1
\label{eq:104}
\end{align}

Notice that $f(1) = 1$. We take the derivative of $f$ with respect to $x$
\begin{align}
f'(x) = \left( \tau_i+1 \right) \left( \dfrac{a + \tau_i x}{a + \tau_i} \right)^{\tau_i+1} \dfrac{\tau_i}{a + \tau_i} x^{-\tau_i} - \left( \dfrac{a + \tau_i x}{a + \tau_i} \right)^{\tau_i+1} \tau_i x^{-\tau_i-1} = 
\tau_i x^{-\tau_i-1} \left( \dfrac{a + \tau_i x}{a + \tau_i} \right)^{\tau_i} \left( \dfrac{x-a}{a + \tau_i} \right)
\label{eq:105}
\end{align}

$f'(x) \geq 0 \Leftrightarrow x \geq a \Leftrightarrow \frac{\hat{\tau}_i}{\tau_i} \geq n + \sum_{k \neq i} t_k$.

$f(\frac{\hat{\tau}_i}{\tau_i})$ is decreasing in $\hat{\tau}_i$ until $\hat{\tau}_i = \tau_i \left( n + \sum_{k \neq i} t_k \right) $, then it is increasing.
This implies that for  $\hat{\tau}_i < \tau_i$  Eq. (\ref{eq:102}) is satisfied, i.e., user $i$ has no incentive 
to report a lower type.
However, if $\hat{\tau}_i \rightarrow \tau_i^+$, since $f'(1)<0$, then user $i$ ha an incentive to communicate a higher 
type (this result is linked to \ref{prop:1}).
In fact Eq. (\ref{eq:102}) is unsatisfied $\forall t_{-i} \in T_{-i}$, and therefore Eq. (\ref{eq:101}) is unsatisfied.
Since the function $f(\frac{\hat{\tau}_i}{\tau_i})$ increases for  $\hat{\tau}_i > \tau_i \left( n + \sum_{k \neq i} t_k \right) $, the only way for Eq. (\ref{eq:102}) to be satisfied is that the function $f(x)$ will eventually reach the value $1$ for a value $x^{th} = \frac{\tau^{th}}{\tau_i}$ and all the types higher than $\tau_i$ are higher than the threshold value $\tau^{th}$. 
Notice that it is sufficient that this condition is verified by the type that follows $\tau_i$, i.e., $\tau_{i+1}$.
Substituting $\hat{\tau}_i$ with $\tau_{i+1}$ into Eq. (\ref{eq:103}) we obtain Eq. (\ref{eq:dis_suf}).

\end{IEEEproof}

\section{Proof of Proposition \ref{prop:a_priori}}\label{app:a_priori}

\begin{IEEEproof}

First, we demonstrate that Eq. (\ref{eq:geo_opt}) describes a convex problem if $\tau_m \leq n$.
The constraints describe a convex set.
We can rewrite the objective function in the following way
\begin{align}
f(d) &= - \ln \left[ \left( \mu - \sum_{i=1}^n d_i \right) \sum_{t \in \mathcal{T}} P_t(t) \prod_{i=1}^n d_i^{\frac{t_i}{n}} \right]
= - \ln \left[ \left( \mu - \sum_{i=1}^n d_i \right) \prod_{i=1}^n \sum_{l=1}^m P(\tau_l) d_i^{\frac{\tau_l}{n}} \right] = \nonumber\\
&= - \ln \left( \mu - \sum_{i=1}^n d_i \right) - \sum_{i=1}^n \ln \sum_{l=1}^m P(\tau_l) d_i^{\frac{\tau_l}{n}}
\label{eq:geo_opt2}
\end{align}

We calculate the partial derivatives of $f(d)$
\begin{align}
&\dfrac{\partial f(d)}{\partial d_j} = \dfrac{1}{\mu - \sum_{i=1}^n d_i} - \dfrac{\sum_{l=1}^m P(\tau_l) \frac{\tau_l}{n} d_i^{\frac{\tau_l}{n}-1}}{\sum_{l=1}^m P(\tau_l) d_i^{\frac{\tau_l}{n}}} \nonumber\\ 
&\dfrac{\partial^2 f(d)}{\partial d_j^2} = \dfrac{1}{\left( \mu - \sum_{i=1}^n d_i \right)^2} - \dfrac{ \left( \sum_{l=1}^m P(\tau_l) \frac{\tau_l}{n} \left( \frac{\tau_l}{n} - 1 \right) d_i^{\frac{\tau_l}{n}-2} \right) \left( \sum_{l=1}^m P(\tau_l) d_i^{\frac{\tau_l}{n}} \right) - \left( \sum_{l=1}^m P(\tau_l) \frac{\tau_l}{n} d_i^{\frac{\tau_l}{n}-1} \right)^2 }{ \left( \sum_{l=1}^m P(\tau_l) d_i^{\frac{\tau_l}{n}} \right)^2 } \nonumber\\ 
&\dfrac{\partial^2 f(d)}{\partial d_j \partial d_k} = \dfrac{1}{\left( \mu - \sum_{i=1}^n d_i \right)^2} 
\label{eq:geo_opt3}
\end{align}

We have $\dfrac{\partial^2 f(d)}{\partial d_j^2} \geq \dfrac{\partial^2 f(d)}{\partial d_j \partial d_k} \geq 0$, where the first inequality is valid if $\tau_m \leq n$.

Before concluding, we state and prove the following Lemma.

\begin{lemma}
The matrix
\begin{equation}
H = \left[ \begin{array}{cccc} 
\alpha_1 & \beta & \ldots & \beta \\ 
\beta & \alpha_2 & \ldots & \beta \\ 
\vdots &  & \ddots & \vdots \\ 
\beta & \beta & \ldots & \alpha_n \\
\end{array}\right] 
\end{equation}
where $\alpha_i \geq \beta \geq 0$, $\forall i = \left\lbrace 1, 2, \cdots, n\right\rbrace $, is positive semidefinite. 
If the first inequality is strict, it is also positive definite.
\label{Hessian}
\end{lemma}

\begin{IEEEproof}
\begin{equation}
H = \beta \left[ \begin{array}{cccc} 
1 & 1 & \ldots & 1 \\ 
1 & 1 & \ldots & 1 \\ 
\vdots &  & \ddots & \vdots \\ 
1 & 1 & \ldots & 1 \\
\end{array}\right] + 
\left[ \begin{array}{cccc} 
\alpha_1-\beta & 0 & \ldots & 0 \\ 
0 & \alpha_2-\beta & \ldots & 0 \\ 
\vdots &  & \ddots & \vdots \\ 
0 & 0 & \ldots & \alpha_n-\beta \\
\end{array}\right]
\end{equation}

Therefore 
\begin{equation}
v^T \cdot H \cdot v = \left( \alpha_1-\beta \right) v_1^2 + \cdots + \left( \alpha_n-\beta \right) v_n^2 + \beta \left( \sum_{i=1}^n v_i \right)^2
\end{equation}
$v^T \cdot H \cdot v \geq 0 \; \forall \, v $ if $\alpha_i \geq \beta \geq 0$ $\forall \, i$.
$v^T \cdot H \cdot v > 0 \; \forall \, v \neq 0$ if $\alpha_i > \beta \geq 0$ $\forall \, i$.
\end{IEEEproof}

Applying Lemma \ref{Hessian} to the Hessian of the function $f(d)$ we obtain that the Hessian is positive 
semidefinite, therefore the function $f(d)$ is convex.

As for the optimality of the a priori incentive compatible mechanism $\left( T, D, \overline{d}, \pi \right)$, for every a priori incentive compatible mechanism we have
\begin{align}
&max_{\pi, d } V_0(d) = max_{\pi, d } \mathbb{E}_t \left[ \mathbb{E}_f \left[ \sqrt[n]{\prod_{i=1}^n U_i^+(f, d, t)} \right] \right] \leq max_{d} \mathbb{E}_t \left[ \sqrt[n]{\prod_{i=1}^n U_i^+(d_0^*, d, t)} \right] = \nonumber\\
&= \max_d \left( \mu - \sum_{i=1}^n d_i \right)^+ \mathbb{E}_t \left[ \sqrt[n]{\prod_{i=1}^n d_i^{t_i}} \right] =
\max_d \left( \mu - \sum_{i=1}^n d_i \right) \mathbb{E}_t \left[ \prod_{i=1}^n d_i^{\frac{t_i}{n}} \right]  
\label{eq:V_0_geo_opt2}
\end{align}
Thus if a randomized intervention rule $\pi$ sustains without intervention $\overline{d}$, the mechanism $\left( T, D, \overline{d}, \pi \right)$ is an optimal a priori incentive compatible direct mechanism.

\end{IEEEproof}

\bibliographystyle{IEEEtran}
\bibliography{IEEEabrv,bibcok}

\begin{thebibliography}{10}
\providecommand{\url}[1]{#1}
\csname url@samestyle\endcsname
\providecommand{\newblock}{\relax}
\providecommand{\bibinfo}[2]{#2}
\providecommand{\BIBentrySTDinterwordspacing}{\spaceskip=0pt\relax}
\providecommand{\BIBentryALTinterwordstretchfactor}{4}
\providecommand{\BIBentryALTinterwordspacing}{\spaceskip=\fontdimen2\font plus
\BIBentryALTinterwordstretchfactor\fontdimen3\font minus
  \fontdimen4\font\relax}
\providecommand{\BIBforeignlanguage}[2]{{%
\expandafter\ifx\csname l@#1\endcsname\relax
\typeout{** WARNING: IEEEtran.bst: No hyphenation pattern has been}%
\typeout{** loaded for the language `#1'. Using the pattern for}%
\typeout{** the default language instead.}%
\else
\language=\csname l@#1\endcsname
\fi
#2}}
\providecommand{\BIBdecl}{\relax}
\BIBdecl

\bibitem{802}
G.~Tan and J.~Guttag, ``The 802.11 {MAC} protocol leads to inefficient
  equilibria,'' in \emph{Proc. IEEE INFOCOM}, vol.~1, 2005, pp. 1--11.

\bibitem{slotAloha}
R.~T.~B. Ma, V.~Misra, and D.~Rubenstein, ``Modeling and analysis of
  generalized slotted-aloha {MAC} protocols in cooperative, competitive and
  adversarial environments,'' in \emph{Proc. IEEE ICDCS}, 2006.

\bibitem{HuangBerry_JSAC06}
J.~Huang, R.~A. Berry, and M.~L. Honig, ``Distributed interference compensation
  for wireless networks,'' \emph{{IEEE} J. Sel. Areas Commun.}, vol.~24, no.~5,
  pp. 1074--1084, 2006.

\bibitem{XiaoMihaela_JSTSP}
Y.~Xiao, J.~Park, and M.~van~der Schaar, ``Intervention in power control games
  with selfish users,'' \emph{IEEE J. Sel. Topics Signal Process., Special
  issue on Game Theory in Signal Processing}, vol.~6, pp. 165--179, 2012.

\bibitem{ParkMihaela_EURASIP}
J.~Park and M.~van~der Schaar, ``Stackelberg contention games in multiuser
  networks,'' \emph{EURASIP Journal on Advances in Signal Processing}, pp.
  1--15, 2009.

\bibitem{YiMihaela_TCOM}
Y.~Su and M.~van~der Schaar, ``Linearly coupled communication games,''
  \emph{{IEEE} Trans. Commun.}, vol.~59, pp. 2543--2553, 2011.

\bibitem{GaiKrishnamachari_Infocom}
Y.~Gai, H.~Liu, and B.~Krishnamachari, ``A packet dropping-based incentive
  mechanism for {M/M/1} queues with selfish users,'' in \emph{Proc. IEEE
  INFOCOM}, 2011, pp. 2687--2695.

\bibitem{BharathKumarJaffe}
K.~Bharath-Kumar and J.~M. Jaffe, ``A new approach to performance-oriented flow
  control,'' \emph{{IEEE} Trans. Commun.}, vol.~29, pp. 427--435, 1981.

\bibitem{DouligerisMazumdar}
C.~Douligeris and R.~Mazumdar, ``A game theoretic perspective to flow control
  in telecommunication networks,'' \emph{Journal of the Franklin Institute},
  vol. 329, no.~2, pp. 383--402, 1992.

\bibitem{ZhangDouligeris_TCOM}
Z.~Zhang and C.~Douligeris, ``Convergence of synchronous and asynchronous
  greedy algorithm in a multiclass telecommunications environment,''
  \emph{{IEEE} Trans. Commun.}, vol.~40, no.~8, pp. 1277--1281, 1992.

\bibitem{Hurwicz}
L.~Hurwicz, ``Game theory,'' \emph{Decision and Organization: a Volume in Honor
  of Jacob Marshak}, pp. 297--336, 1972.

\bibitem{DsHamMaskin}
P.~Dasgupta, P.~Hammond, and E.~Maskin, ``The implementation of social choice
  rules: some results on incentive compatibility,'' \emph{Review of Economic
  Studies}, vol.~46, pp. 185--216, 1979.

\bibitem{Holmstrom}
B.~Holmstrom, ``Moral hazard and observability,'' \emph{Bell Journal of
  Economics}, vol.~10, pp. 74--91, 1979.

\bibitem{Myerson1979}
R.~B. Myerson, ``Incentive-compatibility and the bargaining problem,''
  \emph{Econometrica}, vol.~47, pp. 61--73, 1979.

\bibitem{Myerson1981}
------, ``Optimal auction design,'' \emph{Mathematics of Operations Research},
  vol.~6, pp. 58--73, 1981.

\bibitem{Myerson1982}
------, ``Optimal coordination mechanism in generalized principal-agent
  problems,'' \emph{Journal of Mathematical Economics}, vol.~10, pp. 67--81,
  1982.

\bibitem{ParkMihaela_JSAC}
J.~Park and M.~van~der Schaar, ``The theory of intervention games for resource
  sharing in wireless communications,'' \emph{{IEEE} J. Sel. Areas Commun.},
  vol.~30, no.~1, pp. 165--175, 2012.

\bibitem{ParkMihaela_Gamenets}
------, ``Designing incentive schemes based on intervention: The case of
  imperfect monitoring,'' in \emph{Proc. GameNets}, 2011.

\bibitem{WangComaniciu_WPC}
D.~Wang, C.~Comaniciu, and U.~Tureli, ``Cooperation and fairness for slotted
  aloha,'' \emph{Wireless Personal Communications}, vol.~43, no.~1, pp. 13--27,
  2007.

\bibitem{YangKimZhangChiangTan_INFOCOM11}
L.~Yang, H.~Kim, J.~Zhang, M.~Chiang, and C.~W. Tan, ``Pricing-based spectrum
  access control in cognitive radio networks with random access,'' in
  \emph{Proc. IEEE INFOCOM}, 2011, pp. 2228--2236.

\bibitem{BasarSrikant02}
T.~Basar and R.~Srikant, ``Revenue-maximizing pricing and capacity expansion in
  a many-users regime,'' in \emph{Proc. IEEE INFOCOM}, 2002, pp. 1556--1563.

\bibitem{ShenBasar07}
H.~Shen and T.~Basar, ``Optimal nonlinear pricing for a monopolistic network
  service provider with complete and incomplete information,'' \emph{{IEEE} J.
  Sel. Areas Commun.}, vol.~25, pp. 1216--1223, 2007.

\bibitem{SchmidtBerry09}
D.~A. Schmidt, C.~Shi, R.~A. Berry, M.~L. Honig, and W.~Utschick, ``Distributed
  resource allocation schemes,'' \emph{IEEE Signal Processing Magazine},
  vol.~26, no.~5, pp. 53--63, 2009.

\bibitem{AlpcanBasar_CDMA}
T.~Alpcan, T.~Basar, R.~Srikant, and E.~Altman, ``{CDMA} uplink power control
  as a noncooperative game,'' \emph{Wireless Networks}, vol.~8, pp. 659--670,
  2002.

\bibitem{HuangBerry_06_Auction}
J.~Huang, R.~A. Berry, and M.~L. Honig, ``Auction-based spectrum sharing,''
  \emph{Mobile Networks and Applications}, vol.~11, pp. 405--418, 2006.

\bibitem{AlpcanBocheNaik}
T.~Alpcan, H.~Boche, and S.~Naik, ``A unified mechanism design framework for
  networked systems,'' \emph{CoRR}, vol. abs/1009.0377, 2010.

\bibitem{PavanJenTara}
P.~Nuggehalli, J.~Price, and T.~Javidi, ``Pricing and {QoS} in wireless random
  access networks,'' in \emph{Proc. IEEE GLOBECOM}, 2008, pp. 1--5.

\bibitem{Harsanyi}
J.~C. Harsanyi, ``Games with incomplete information played by `bayesian'
  players,'' \emph{Management Science}, vol.~14, no.~5, pp. 320--334, 1968.

\bibitem{Giessler}
A.~Giessler, J.~Hanle, A.~Konig, and E.~Pade, ``Free buffer allocation - an
  investigation by simulation,'' \emph{Comput. Networks}, vol.~1, pp. 199--204,
  1978.

\bibitem{Kleinrock}
L.~Kleinrock, ``Power and deterministic rules of thumb for probabilistic
  problems in computer communications,'' \emph{Conference Record, International
  Conference on Communications}, vol.~2, no.~4, pp. 43.1.1--43.1.10, 1979.

\bibitem{Holmstrom1980}
\BIBentryALTinterwordspacing
B.~Holmstrom, ``On the theory of delegation,'' Northwestern University, Center
  for Mathematical Studies in Economics and Management Science, Discussion
  Papers 438, Jun. 1980. [Online]. Available:
  \url{http://ideas.repec.org/p/nwu/cmsems/438.html}
\BIBentrySTDinterwordspacing

\bibitem{CrawSobel}
V.~P. Crawford and J.~Sobel, ``Strategic information transmission,''
  \emph{Econometrica}, vol.~50, no.~6, pp. 1431--1451, 1982.

\bibitem{Kakutani}
S.~Kakutani, ``A generalization of {Brouwer}'s fixed point theorem,''
  \emph{Duke Mathematical Journal}, vol.~8, no.~3, pp. 457--459, 1941.

\end{thebibliography}

\end{document}